\documentclass[preprint2,english]{aastex}

\shorttitle{}
\shortauthors{K. Lakhchaura et al.}
\usepackage{psfig}
\usepackage{epsfig}
\usepackage{rotating}
\usepackage{graphicx}
\usepackage{rotate}
\usepackage{subfigure}
\usepackage{threeparttable}
\usepackage[abs]{overpic}
\usepackage[dvips]{color}
\begin{document}
\title{A CLUSTER PAIR : A3532 AND A3530}

\author{Kiran Lakhchaura\altaffilmark{1} and K. P. Singh}
\affil{Department of Astronomy and Astrophysics, Tata Institute of Fundamental Research, \\1 Homi Bhabha Road, Mumbai 400 005, India}

\author{D. J. Saikia}
\affil{National Centre for Radio Astrophysics, Tata Institute of Fundamental Research, Pune University Campus, Pune 411 007, India}

\and

\author{R. W. Hunstead}
\affil{Sydney Institute for Astronomy, School of Physics, University of Sydney, NSW 2006, Australia}

\altaffiltext{1}{e-mail : kiran\_astro@tifr.res.in}

\begin{abstract}

We present a detailed study of a close pair of clusters of galaxies, A3532 and A3530, and their environments. The \textit{Chandra} X-ray image 
of A3532 reveals presence of substructures on scales of $\sim$20$^{\prime\prime}$ in its core. XMM-Newton maps of the clusters show excess X-ray 
emission from an overlapping region between them. Spectrally determined projected temperature and entropy maps do not show any
 signs of cluster scale mergers either in the overlapping region or  in any of the clusters. In A3532, however, some signs of the presence of 
galaxy scale mergers are visible e.g., anisotropic temperature variations in the projected thermodynamic maps, a wide angled tailed (WAT) radio
 source in the brighter nucleus of its dumbbell Brightest Cluster Galaxy (BCG), and a candidate X-ray cavity coincident with the northwestern 
extension of the WAT source in the low-frequency radio observations. The northwestern extension in A3532 seems either a part of the WAT or an 
unrelated diffuse source in A3532 or in the background. There is an indication that the cool core in A3532 has been disrupted by the central 
AGN activity. A reanalysis of the 
redshift data reinforces the close proximity of the clusters. The excess emission in the overlapping region appears to be a result of tidal 
interactions as the two clusters approach each other for the first time. However, we can not rule out the possibility of the excess being due to 
the chance superposition of their X-ray halos.
\end{abstract}


\section{Introduction}
\label{sec:Intro}
 Clusters of galaxies are believed to form hierarchically by 
the merger of smaller groups and clusters. Major cluster mergers, which are believed to be the most energetic events in the Universe 
(\cite{Sarazin:2002}), involve two clusters of similar masses (\cite{PlanellQuil:2009}). Many clusters are often found to form large 
concentrations, called superclusters (\cite{Shapley:1933}), which are the largest (size$\sim10$-$100$ Mpc) systems of galaxies known to us 
(\cite{Vogeley:1994}). 
Rich superclusters are also appropriate systems for studying major cluster mergers. The galaxy number density in superclusters is $\sim$10 
times that in the field on length scales of $\sim$10h$^{-1}_{100}$ Mpc (\cite{Bard:2000}). This results in high peculiar velocities of 
the galaxies in superclusters, which increases the probability of cluster$-$cluster collisions (\cite{Bard:2001}). One such example of 
a rich supercluster environment is the central region of the Shapley Supercluster (SSC). \cite{Raychaudhury:1991}, by using X-ray 
observations from \textit{Einstein Observatory} and \textit{EXOSAT}, and optical data from Automatic Plate Measuring (APM) facility at 
Cambridge, studied 17 clusters from the core region of the SSC. The X-ray data showed an exceptionally high density of rich clusters 
($\sim$10-50) and multiple X-ray peaks in most of the clusters belonging to this region. By using the optical data they estimated an 
exceptionally large mass (between $1.4\times10^{16}$ and $1.6\times10^{17}$ h$^{-1}_{50}$ $M_{\odot}$) for the 
\textquoteleft{}core\textquoteright{} region (diameter 74 h$^{-1}_{50}$ Mpc) of the SSC. The optical data also revealed a large deviation from 
the Hubble flow within SSC which suggested that it might be a nearly bound system. By using the X-ray observations from the ROSAT All Sky 
Survey (RASS), \cite{deFilippis:2005} discovered 14 new cluster candidates, and also observed that the over-density of clusters in the SSC 
outskirts is mainly due to an excess of low X-ray luminosity clusters. This led to a suggestion that the whole region is still accreting small 
clusters with low luminosities from the outskirts (\cite{deFilippis:2005}). 

The SSC hosts two major cluster complexes dominated by A3558 and A3528. This concentration of clusters is a rich system for observing cluster 
mergers at various evolutionary stages (\cite{Bard:2001}). The A3558 complex comprises 
three rich clusters of galaxies viz., A3556, A3558, and A3562, and several other poor clusters and groups, while the A3528 complex is formed by
 the galaxy clusters A3528, A3532 and A3530. The properties of the A3558 complex have been extensively studied using multiwavelength studies 
(optical: \cite{Bard:1994,Bard:1998a,Bard:1998b}; radio: \cite{Venturi:1997,Venturi:1998,Venturi:2000}; X-ray: \cite{Bard:1996,Ettori:2000}). The 
less well studied A3528 cluster complex has a mean redshift of $\overline{z}$ = $0.0535$ and is elongated in the north-south direction. Using 
ROSAT Position Sensitive Proportional Counter (PSPC) observations, \cite{Schindler:1996} found that A3528 is a double cluster comprising A3528N 
and A3528S. Using a redshift survey of galaxies, \cite{Bard:2001} detected substructures in the A3528 complex by using the non-parametric
 and scale-independent DEDICA method (\cite{Pisani:1993,Pisani:1996}) in both bi-dimensional and three-dimensional samples. From the bi-dimensional sample
 the cluster A3528 was found to contain a total of ten groups, two of which appear to be associated with A3528N and A3528S. From the 
three-dimensional sample, the A3532-A3530 system was found to contain five groups, out of which two seem to be associated with A3530 and A3532,
 while two others are found at the intersection of the Abell radii of the two clusters, with a mean velocity in agreement with the main 
components. The cluster complex has been studied in the radio by \cite{Venturi:2001} using 13-cm and 22-cm observations carried out with the 
Australia Telescope Compact Array (ATCA). \cite{MaudMam:2007} studied the radio continuum emission from galaxies in the SSC core region and
 found that the galaxies in the A3528 complex are marginally more radio-luminous than elsewhere, contrary to what is observed in the 
neighboring A3558 cluster complex, where the galaxies have lower radio luminosity and radio loudness compared to the field galaxies. They have 
attributed the decrease in the radio loudness in A3558 cluster complex to starvation of the AGN in them, and reduced star formation 
activity due to the enhanced ram-pressure stripping in merging clusters. The lack of decrease in the radio loudness for the A3528 cluster 
complex is, therefore, probably because in this region the clusters are approaching for the first time (\cite{MaudMam:2007}). 
In this paper, we present a detailed study of the environments, interactions and internal dynamics 
of the two less well studied clusters of galaxies viz., A3530 and A3532, from the A3528 cluster complex, using mainly radio and X-ray 
observations.

  The paper is organized as follows. Detailed information about the two clusters and a summary of their previous X-ray, radio and optical 
observations are given in \S\ref{sec:the_a3530_a3532_system}. Details of the X-ray and radio observations used here, along with the data 
reduction, are presented in \S\ref{sec:Obs_n_data_red}. The X-ray, optical and radio morphologies of the clusters 
are described in \S\ref{sec:X-ray_morphology}. Results from the global spectral analyses and the X-ray luminosity estimates are provided in 
\S\ref{sec:Total-Spec-Analy} and \S\ref{sec:luminosity_estimates}, respectively. Results from the azimuthally averaged (projected and 
deprojected) spectral analyses and the two-dimensional projected thermodynamic maps of the clusters are described in 
\S\ref{sec:Azimuth_spec_analys} and \S\ref{sec:box_thermodynamic_maps}, respectively. The estimates of the cooling time, gas mass and virial 
mass for the two clusters are given in \S\ref{sec:cool_time_estimate}, \S\ref{sec:gas_mass_estimates}, and 
\S\ref{sec:gal_vel_distrb_an_vir_mas}, respectively. A discussion based on the results is in \S\ref{sec:discussion}. A lambda cold dark 
matter cosmology with $H_{0}$ = 70 km s$^{-1}$ Mpc$^{-1}$ and $\Omega_{M}$ = 0.3 ($\Omega_{\Lambda}$ = 0.7) has been assumed throughout.

\section{The A3530-A3532 System}
\label{sec:the_a3530_a3532_system}
Both A3532 and A3530 are regular clusters of richness class 0, with 36 and 34 member galaxies, respectively (\cite{AbCoOl:1989}). The 
positional coordinates (R.A.(J2000), Decl.(J2000)) of A3532 are $12^{h} 57^{m} 19.2^{s} $, $-30^{d} 22^{\prime} 13^{\prime\prime}$ 
(\cite{MahGe:2001}), and those of A3530 are $12^{h} 55^{m} 36^{s} $, $-30^{d} 21.2^{\prime}$ (Abell et al. 1989). A3532 and A3530 are 
located at mean redshifts of $\overline{z}=0.0554\pm0.0004$ (\cite{Cristiani:1987}) and $\overline{z}=0.0543\pm0.0009$ (\cite{Vettolani:1990}),
 respectively. The total mass (M$_{500}$) (sum of baryonic and non-baryonic mass within R$_{500}$, i.e., radius within which the mean density 
of the cluster equals 500 times the critical density) estimates of A3532 and A3530, reported by \cite{Ettori:1997}, based on an image 
deprojection analysis done using \textit{ROSAT} PSPC images, are $6.38\times10^{14}$ h$^{-1}_{50}$ M$_{\odot}$ and $4.69\times10^{14}$ 
h$^{-1}_{50}$ M$_{\odot}$, respectively.  
\cite{Bard:2001} reported that the projected distance between the centers of A3530 and A3532 is smaller than the sum of 
their virial  radii, thus indicating a tidal interaction between the two clusters. It should, however, be noted that, if the redshift difference 
between the two clusters ($\vartriangle z \sim 0.0010\pm0.0009$) is entirely due to Hubble flow, then the difference in their line of sight 
distances is 4.2$\pm3.7$ h$^{-1}_{70}$ Mpc, which can be larger than the sum of their virial radii ($\sim$3 h$^{-1}_{70}$ Mpc). This 
means that it is possible that the two clusters do not actually overlap. However, a tidal interaction between them is still possible and they 
may form a bound system.  Various estimates of the temperatures and X-ray luminosities of the two clusters can be found in the literature 
(\cite{Raychaudhury:1991,Ettori:1997,McCarthy:2002,ReipBoh:2002}). However, the results obtained by Ikebe et
 al. (2002) used ASCA with the energy band closest to \textit{Chandra} and \textit{XMM-Newton}. By fitting 2-T thermal plasma models to the 
ASCA spectra of A3532 and A3530, \cite{Ikebe:2002} obtained hot component temperatures of 4.4$\pm$0.2 and 4.1$\pm$0.3 keV, and 0.1-2.4 keV 
X-ray luminosities of $(5.9\pm0.2)\times10^{43}$h$^{-2}_{100}$ ($=(12.0\pm0.4)\times10^{43}$h$^{-2}_{70}$) and 
$(3.3\pm0.2)\times10^{43}$h$^{-2}_{100}$ ($=(6.6\pm0.3)\times10^{43}$h$^{-2}_{70}$) erg cm$^{-2}$ s$^{-1}$, respectively. Note that Ikebe et 
al. have provided the temperature values of the hot components only, as these are the temperatures of the main extended emission components of 
the clusters and, therefore, are good measures of the virial temperatures of the clusters (\cite{Ikebe:2002}). A comparison of the values 
obtained by \cite{Ikebe:2002} with those obtained from our analysis has been made in Table \ref{tab:total_regions_spectral_results}.

 By using ASCA and ROSAT observations, \cite{Chen:2007} identified both A3532 and A3530 as non cool core (NCC) clusters with cooling 
times of $(2.4^{+0.5}_{-0.4})\times10^{10}$ and $(5.3\pm0.6)\times10^{10}$ h$^{-1/2}_{50}$ yr, respectively. The virial radii R$_{200}$ of 
A3532 and A3530, computed by \cite{Vulcani:2011}, using their respective velocity dispersions of $621\pm53$ km s$^{-1}$ and $563\pm52$ km 
s$^{-1}$ ( \cite{Cava:2009}), are 1.50 h$^{-1}_{50}$ Mpc and 1.36 h$^{-1}_{50}$ Mpc, respectively. 

 A3530 is seen to host a pair of  elliptical BCGs at its center (\cite{Venturi:2001}), whereas A3532 is found to be dominated by a 
dumbbell system of BCGs at the center (\cite{Wirth:1982}; \cite{Parma:1991}; \cite{Machacek:2007}; \cite{Gregorini:1994}). \cite{Pimbblet:2008} 
studied a complete sample of dumbbell galaxies in the southern rich clusters listed in \cite{Gregorini:1994}, and found that 
while most of the dumbbell BCGs (including that in A3532) have at least one dumbbell component with a significant peculiar velocity, the 
absence or presence of subclustering in the dumbbell BCG clusters is due to the different states of post merger activity.

 Clusters of galaxies are found to host a wide variety of radio source morphologies, such as the characteristic Wide Angle 
Tailed (WAT) structure (\cite{Owen_Rudnick:1976}; \cite{Roettiger:1996}; \cite{Douglass:2008}; \cite{Mao:2010}). Specially, the dumbbell galaxy 
systems, as seen in A3532, are very often seen to host radio sources in one or both of their members. A Wide-Angle Tail (WAT) radio source is 
indeed associated with the brighter nucleus of the dumbbell BCG in A3532 in the ATCA 13-cm and 22-cm images (Venturi et al. 2001). Close 
proximity of the galaxies ($\sim$10-30 h$^{-1}_{70}$ kpc) in the dumbbell systems can dynamically affect the radio jets through gravitational 
interaction with the confining gas cloud (Wirth et al. 1982). Here, it should be noted that the separation between the dumbbell galaxies in 
A3532 is only $\sim$25 h$^{-1}_{70}$ kpc in projection. The radio sources associated with the dumbbells are often found to be different from 
single galaxy radio sources because of their more distorted (irregular) structures, and flatter radio luminosity functions, indicating a 
triggering of the radio source in the main galaxy by its close companion (Parma et al. 1991, Gregorini et al. 1994), or alternatively 
causing an increase in the luminosity of the existing radio source (Parma et al. 1991). The total radio flux densities estimated for the WAT 
source in A3532 at 13-cm and 22-cm, as reported by Venturi et al. (2001), are 651.7 mJy and 1056.6 mJy (with errors $\sim$10-15\%), 
respectively. Combining these with the 6-cm flux density of the source as given in Gregorini et al. (1994), Venturi et al. also 
computed the spectral index $\alpha\sim0.85$, for the 6-22 cm wavelength range.

\setcounter{secnumdepth}{4}
\section{Observations and Data Reduction}  
\label{sec:Obs_n_data_red}
  The cluster A3532 has been observed with both \textit{XMM-Newton} and \textit{Chandra}, whereas A3530 has only been observed using 
\textit{XMM-Newton}. A journal of the X-ray observations is given in Table \ref{tab:observation_table}. Throughout 
\S\ref{sec:Obs_n_data_red} and \S\ref{sec:xray_radio_analy_results}, we have adopted an average redshift of 
0.0554 (\cite{Cristiani:1987}) for A3532 and 0.0543 (\cite{Vettolani:1990}) for A3530.
\subsection{X-ray Data} 
\label{sec:xray_data}
\subsubsection{XMM-Newton}  
\label{sec:XMM_data_analysis}

  A3532 and A3530 were observed with \textit{XMM-Newton} on 2002 July 03 and 2004 January 15 respectively (see Table 
\ref{tab:observation_table}). For both the observations the three EPIC cameras MOS1, MOS2 (\cite{Turner:2001}) and 
PN (\cite{Struder:2001}) were operated in the full frame mode with the medium filter. The data were obtained from the HEASARC archives.

  All data analyses have been done following the standard procedures from the Science Analysis System (SAS) software version 11.0.0. 
The good time intervals (GTIs), obtained after the initial filtering of the MOS1, MOS2 and PN observations of A3530, 
are 11.4 ks, 11.5 ks and 10.2 ks, respectively, while for A3532 the GTIs are 7.4 ks and 7.9 ks for the observations done using MOS1 and MOS2, 
respectively. The PN observation of the cluster A3532 was strongly affected by soft proton (SP) flares. Only 4.7 ks (of the total exposure 
time of 16.9 ks) exposure was left after the light curve cleaning, which was still likely to be contaminated with residual 
soft protons. Hence, the PN observations of A3532 were found unusable for the study.

\paragraph{Background Treatment :}  
\label{sec:XMM_Background_treatment}
\vspace{0.5cm}
 The residual soft proton contamination, left after the routine temporal filtering was modeled as described in \cite{Snowden:2008}. The quiescent 
particle-induced background and the cosmic background component were removed by using the blanksky observations described
 in \cite{CartRead:2007}. The blanksky event files were filtered as described in \cite{Lakhchaura:2011}. Local background subtraction could
 not be done for either observation, since the sources fill almost the entire field of view of the detectors and emission-free regions were 
difficult to find.

\paragraph{Point Source Removal and Mosaicking :}  
\label{sec:pnt_src_rmvl_n_mosaic}
\vspace{0.5cm}
  Point sources were detected and removed using the SAS task \textit{cheese}. Each of the detected sources was then checked in the 
MOS detector images of both the clusters and spurious sources (detections that did not look like real sources in the images) were 
removed. Finally a total of 26 sources for the cluster A3532 and 12 sources for the cluster A3530 were confirmed. Images and 
exposure maps were created from the filtered and point-source-removed event files from MOS observations of both the clusters in the 
energy band 0.3-9 keV. From these, the mosaicked and 
exposure corrected image with a pixel bin size of $5^{\prime\prime}$ was created using the SAS task \textit{emosaic}. The final 
contour map of the diffuse X-ray emission after smoothing with a Gaussian kernel of width $35^{\prime\prime}$ is shown as Figure 
\ref{fig:A3532_A3530_combined_MOS_image}. Note that, the galaxy groups found by Bardelli et al. (2001) 
(see \S\ref{sec:Intro}), at the intersections of the Abell radii of A3532 and A3530, have not been covered by either of the two 
\textit{XMM-Newton} pointings. An overlay of the X-ray contours (black) superimposed on an optical image of the two clusters from the 
SuperCOSMOS survey in the B$_{J}$ band is shown in Figure \ref{fig:A3532_A3530_SuperCOSMOS_optical_image}.

\subsubsection{Chandra}
\label{sec:Chandra_data_analysis}
\begin{figure*}
\centering
\includegraphics[width=7.0in]{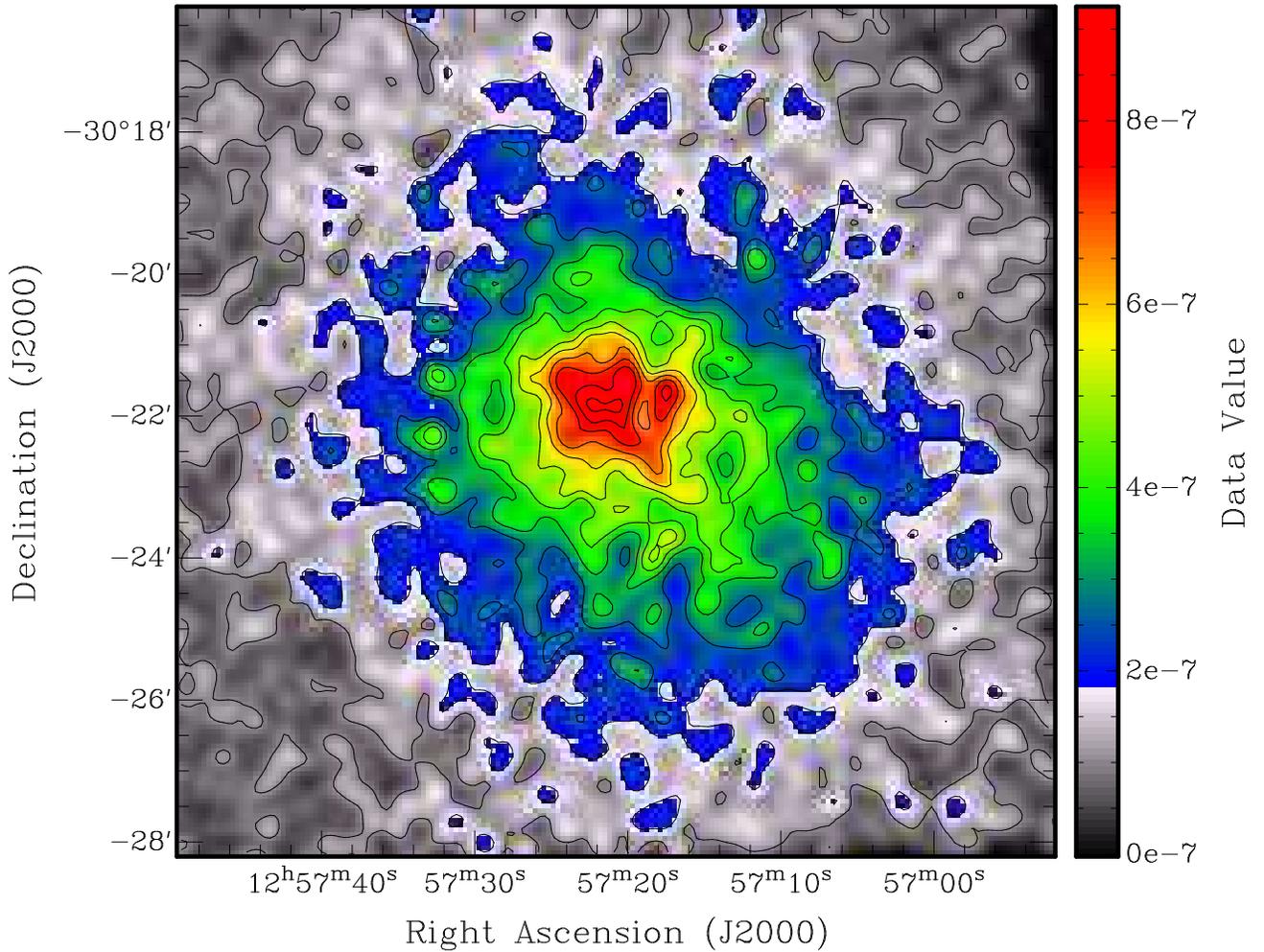}
\caption{Exposure-corrected \textit{Chandra} ACIS image of the A3532 cluster in the 0.3-7.0 keV band (after point source removal and smoothing 
with a Gaussian kernel of width 8$^{\prime\prime}$). The overlaid X-ray emission contours (black) are linearly distributed between  
$1\times10^{-7}$ and $9\times10^{-7}$ counts cm$^{-2}$s$^{-1}$pixel$^{-1}$.}
\label{fig:Chandra_smoothedimage}
\end{figure*}

  A3532 was observed with \textit{Chandra} on 2009 December 2 with ACIS-I detector for 9.8 ks (Table~\ref{tab:observation_table}).
 For the analysis of \textit{Chandra} data we have used CIAO version 4.3 and CALDB version 4.4.0. The data 
were reprocessed using the CIAO script \textit{chandra\_repro}. For all the GTIs, no significant flaring was seen in the 
lightcurves. Point sources were detected using the CIAO task \textit{celldetect}, which were then checked by comparing them with those 
detected with \textit{XMM-Newton} and with the ones seen in the optical image (Fig. \ref{fig:A3532_A3530_SuperCOSMOS_optical_image}). 
 The confirmed point sources were then removed from both the image and the event files and each of the holes created in the image were filled 
with values equal to the average counts in their immediately surrounding pixels, using the ciao task \textit{dmfilth}. The image was then 
exposure corrected and smoothed using a Gaussian kernel of width $8^{\prime\prime}$. The resultant image is shown in Figure 
\ref{fig:Chandra_smoothedimage}.

\subsection{Radio Data}
\label{radio_data}

 We have analyzed archival data from pointed observations of A3532 with the Australia Telescope Compact Array (ATCA), the Giant 
Metrewave Radio Telescope (GMRT), and the Very Large Array (VLA). In addition to the pointed observations, we have used survey data from the 
TIFR (Tata Institute of Fundamental Research) GMRT Sky Survey (TGSS), the NRAO (National Radio Astronomy Observatory) VLA Sky Survey (NVSS), 
the Sydney University Molonglo Sky Survey (SUMSS), the Molonglo cross telescope, and the VLA Low-frequency Sky Survey (VLSS). All details of the 
radio observations are given in Table \ref{tab:radio_observation_table}.

 The ATCA observations were made simultaneously at 1380 and 2378 MHz in multiple-cut mode. The primary flux 
density calibrator was B1934$-$638 and the phase calibrator was B1308$-$220. Data reduction was carried out in MIRIAD (Sault et al. 1995) 
using standard techniques. The GMRT observation at 614 MHz was made with 3C286 as primary flux density calibrator and J1311$-$222 as phase 
calibrator. The VLA observations were made with two IFs, bandwidths of 50 MHz, and center frequencies of 1490 MHz and 4860 MHz. The primary flux 
density calibrator for both VLA observations was 3C286, and the phase calibrators were B1245$-$197 and B1255$-$316. The GMRT and 
VLA data were both processed using the NRAO Astronomical Image Processing System (AIPS) software and standard procedures.
Figures \ref{fig:A3532_VLA_20cm_WAT_source} to \ref{fig:A3532_MOST_36cm_WAT_source} show the radio images produced from the VLA 20 cm, ATCA 
13 cm, GMRT 50 cm, VLA 6 cm, TGSS 2 m, VLSS 4 m, and SUMSS 36 cm observations, respectively.

\section{Analysis and Results}
\label{sec:xray_radio_analy_results}
\subsection{X-ray, Optical and Radio Maps}
\label{sec:X-ray_morphology}
\begin{figure*}
\centering
\includegraphics[width=6.0in]{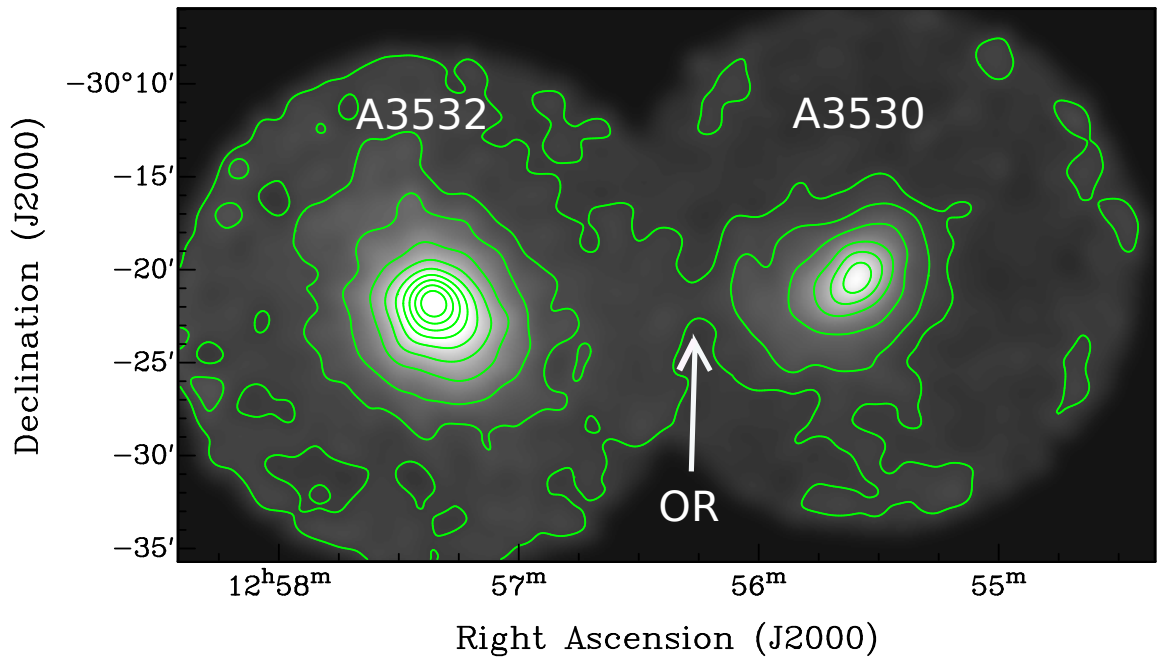}
\caption{The combined and contoured image of A3532 and A3530 clusters from the \textit{XMM-Newton} MOS detectors, smoothed with a Gaussian 
kernel of width $35^{\prime\prime}$, as described in Section \ref{sec:pnt_src_rmvl_n_mosaic}. Contour levels are (1.25, 1.69, 2.50, 3.75, 
5.00, 6.25, 7.50, 8.75, 10.00, 11.25)$\times10^{-8}$ counts s$^{-1}$ arcsec$^{-2}$. Positions of the clusters A3532 and A3530, and their 
overlapping region (OR) are shown. The surface brightness peaks of the two clusters are separated by approximately 1.7 Mpc. 
The outermost contour is about 3 sigma level above the neighboring local background.}
\label{fig:A3532_A3530_combined_MOS_image}
\end{figure*}

  The smoothed and point source removed X-ray image of the two clusters A3530 and A3532 from the \textit{XMM-Newton} MOS (MOS1 + MOS2)
detectors (\S \ref{sec:pnt_src_rmvl_n_mosaic}, Figure \ref{fig:A3532_A3530_combined_MOS_image}) shows that the X-ray emission from A3532 fills 
the entire field of view and is slightly elliptical with the major axis oriented along the NE-SW direction ($\sim45^{\circ}$ to the North). 
X-ray emission from A3530 is much less extended but much more elongated along the NW-SE direction ($\sim$-52$^{\circ}$ to the North). The X-ray
 contours in the inner parts of A3532, seem to be compressed towards the NE and stretched out towards the SW, and those of A3530 seem to be 
stretched out towards the SE. The projected separation between the X-ray peaks of the two clusters is $\sim26.5^{\prime}$, which at their 
average redshift, corresponds to a linear projected separation of $\sim 1.7$ Mpc. The X-ray contours of the two clusters overlap in a 
region common to both clusters (marked in Figure \ref{fig:A3532_A3530_combined_MOS_image}), where they may be tidally interacting with 
each other. Like \textit{XMM-Newton}, the \textit{Chandra} image of A3532 also shows an elongation of the X-ray 
emission in the NE-SW direction with the X-ray contours stretching out more towards the SW, than in the NE direction (Figure 
\ref{fig:Chandra_smoothedimage}). The inner parts of the Chandra image show very disturbed morphology with multiple peaks and subpeaks. This is
 discussed in detail in \S\ref{sec:xray_radio_interaction}.

\begin{figure*}
\centering
\includegraphics[width=7.0in]{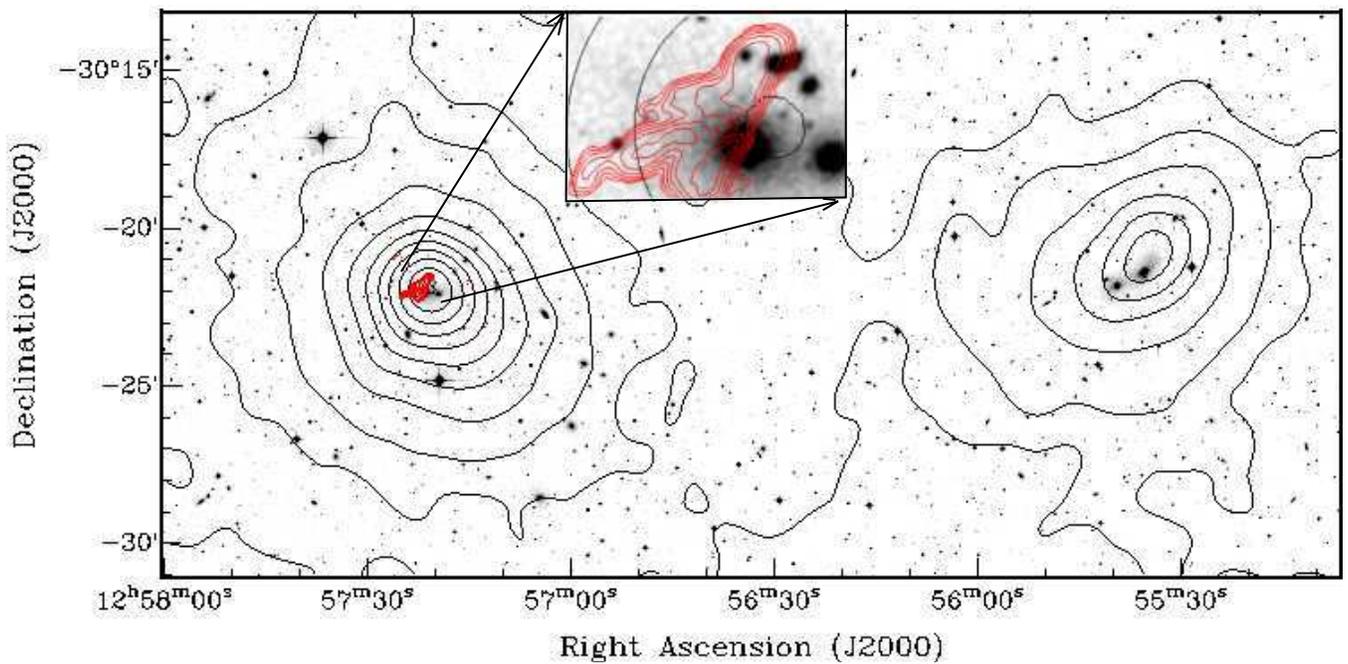}
\caption{Combined image of the clusters A3532 and A3530 from the SuperCOSMOS survey in the $B_J$ band, overlaid with the X-ray  
contours (black) from Figure \ref{fig:A3532_A3530_combined_MOS_image} and also with the VLA 20 cm contours (red, contour levels: -0.3, 
0.3, 0.6, 1.2, 2.4, 4.8 and 9.6 and 19.2 and 38.4 mJy/beam).}
\label{fig:A3532_A3530_SuperCOSMOS_optical_image}
\end{figure*}

\begin{figure*}
\centering
\subfigure
{
\includegraphics[width=2.2in]{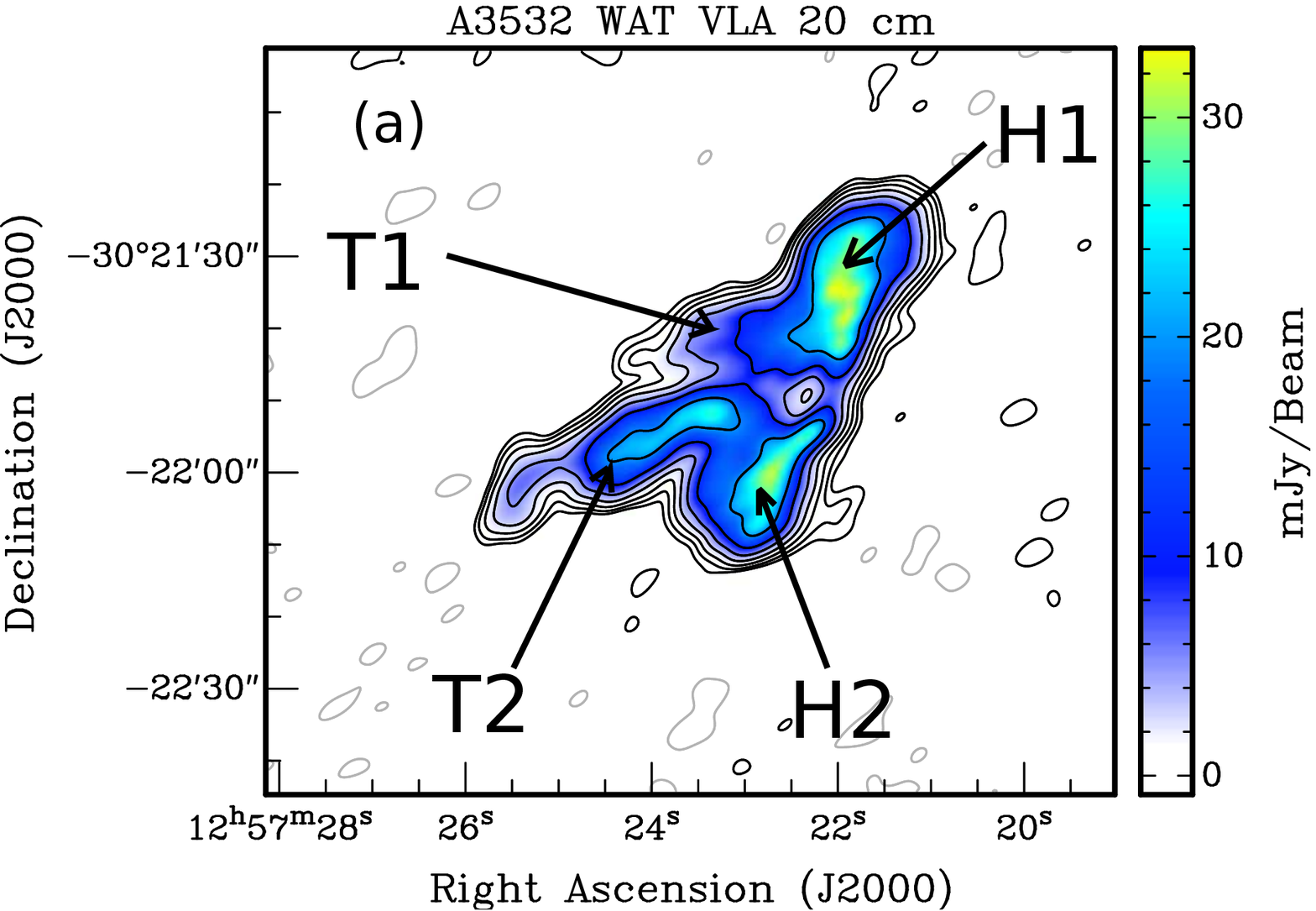}
\label{fig:A3532_VLA_20cm_WAT_source}
}
\subfigure
{
\includegraphics[width=2.2in]{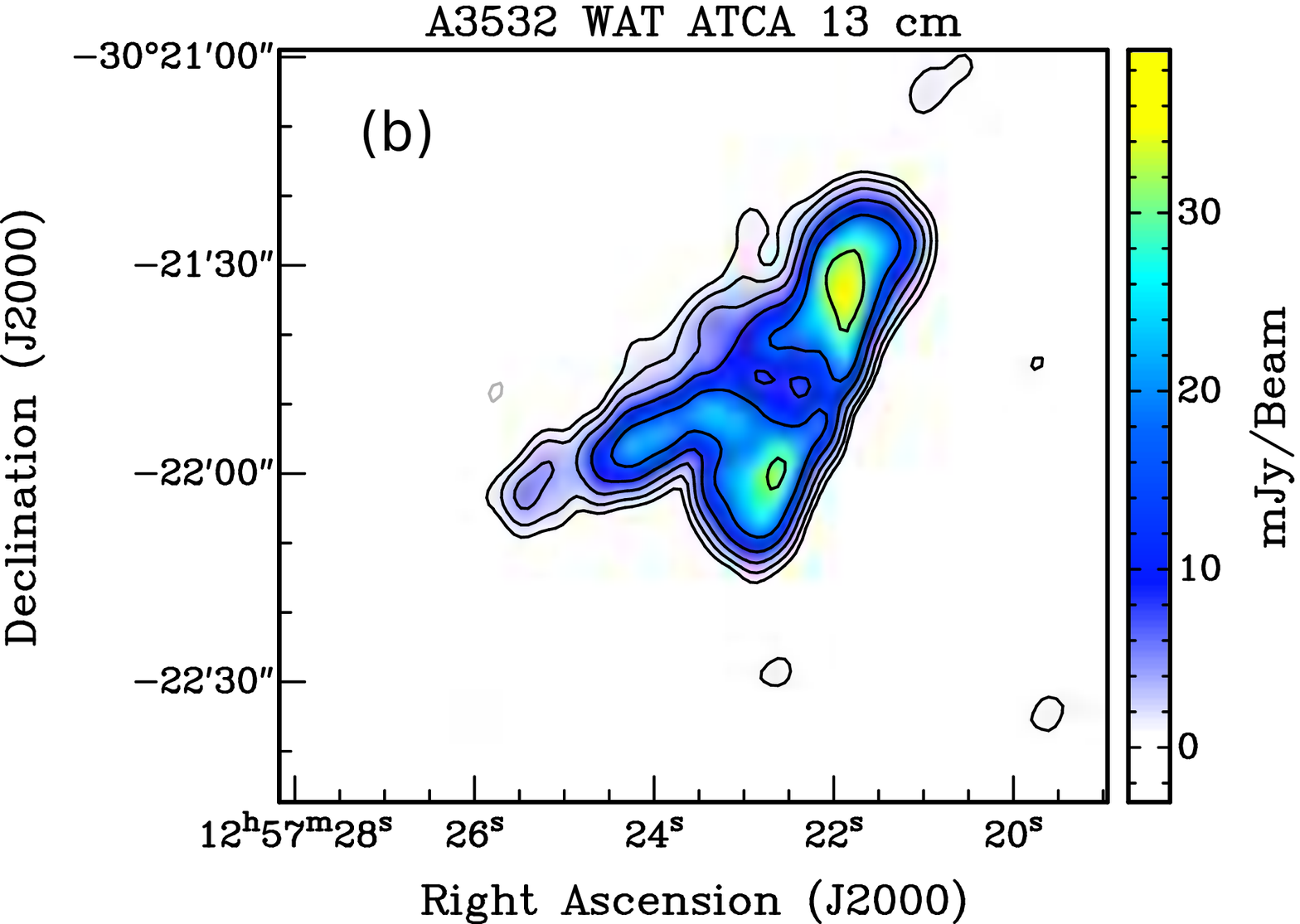}
\label{fig:A3532_atca_13cm_WAT_source}
}\\
\subfigure
{
\includegraphics[width=2.2in]{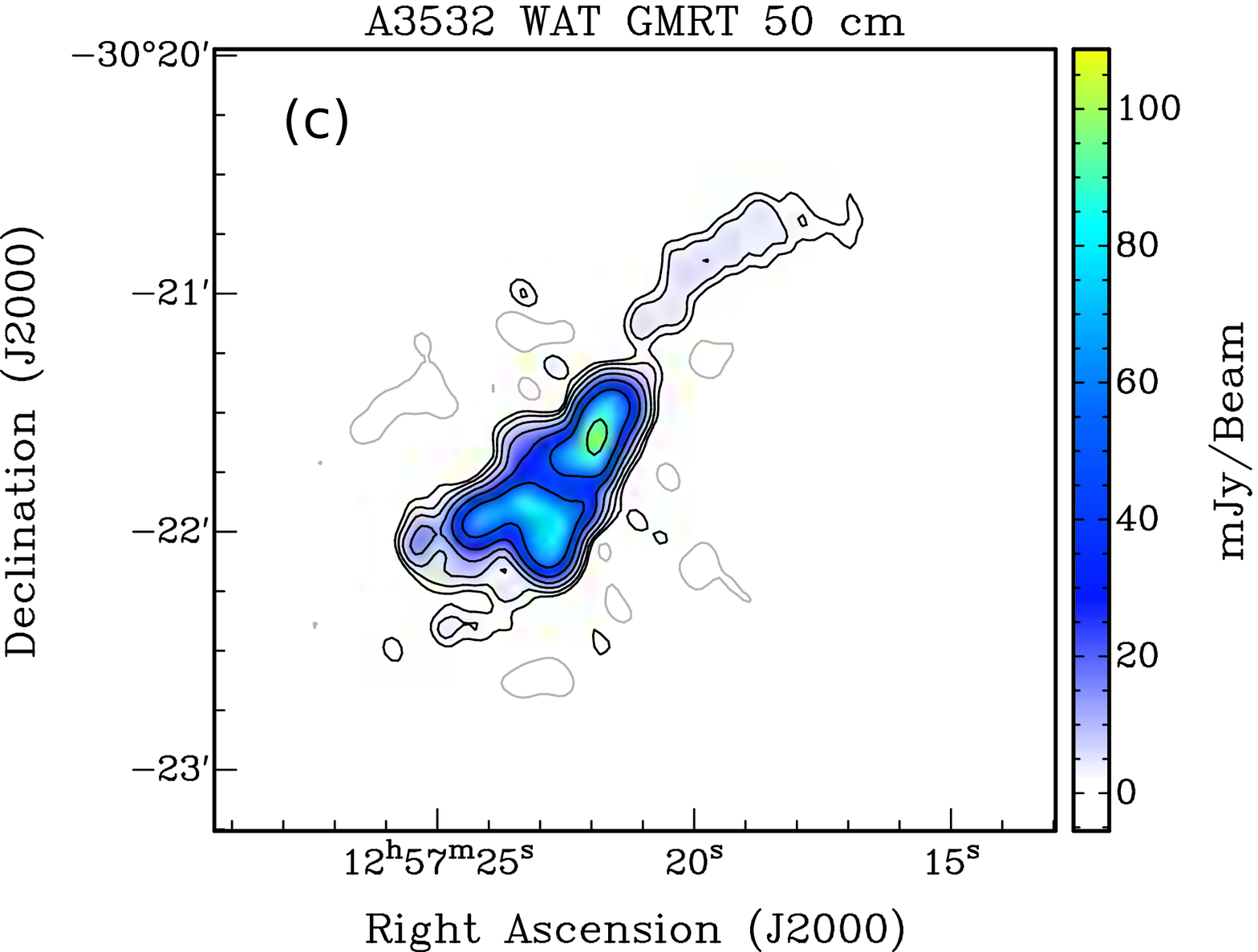}
\label{fig:A3532_GMRT_50cm_WAT_source}
}
\subfigure
{
\includegraphics[width=2.15in]{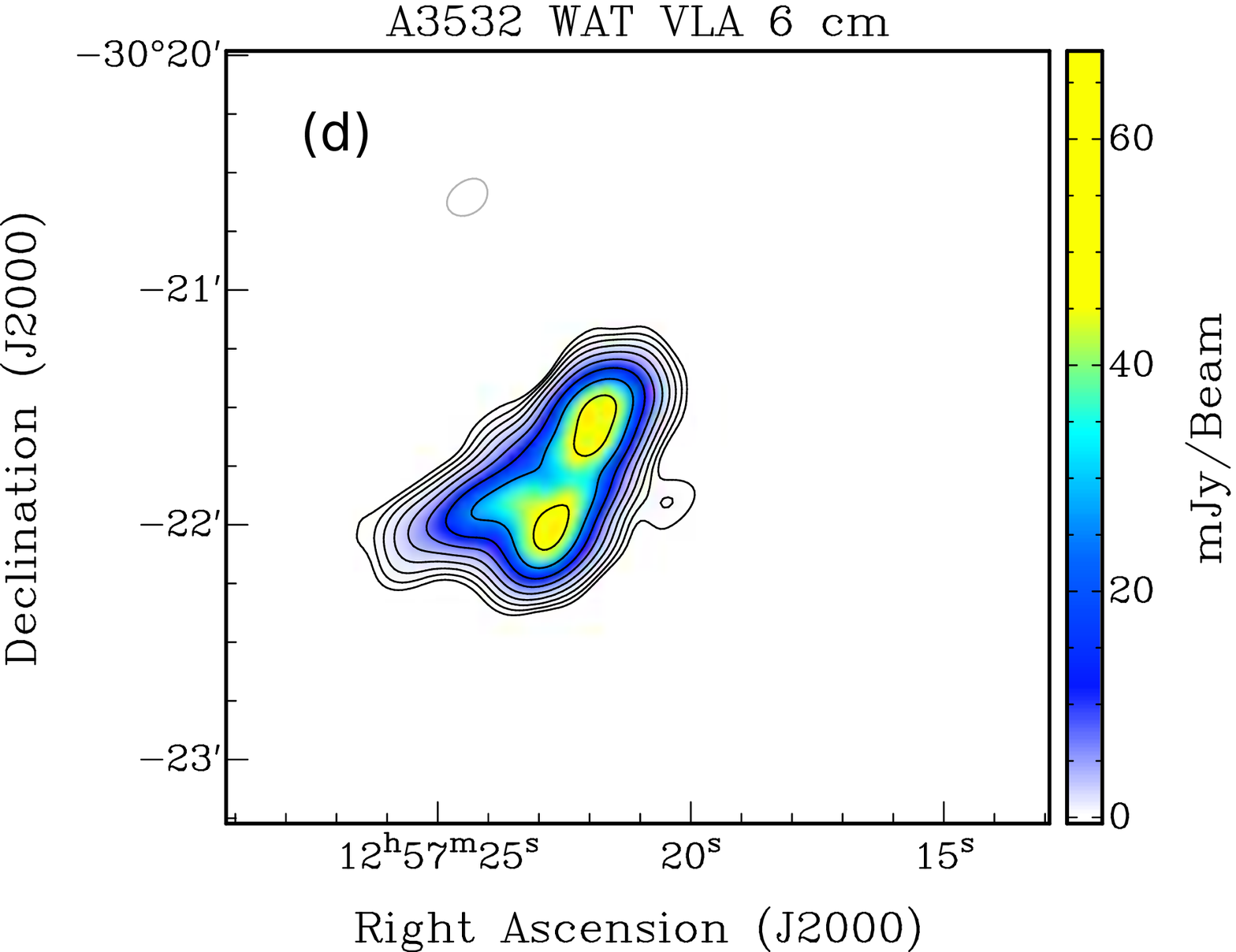}
\label{fig:A3532_VLA_6cm_WAT_source}
}\\
\subfigure
{
\includegraphics[width=2.2in]{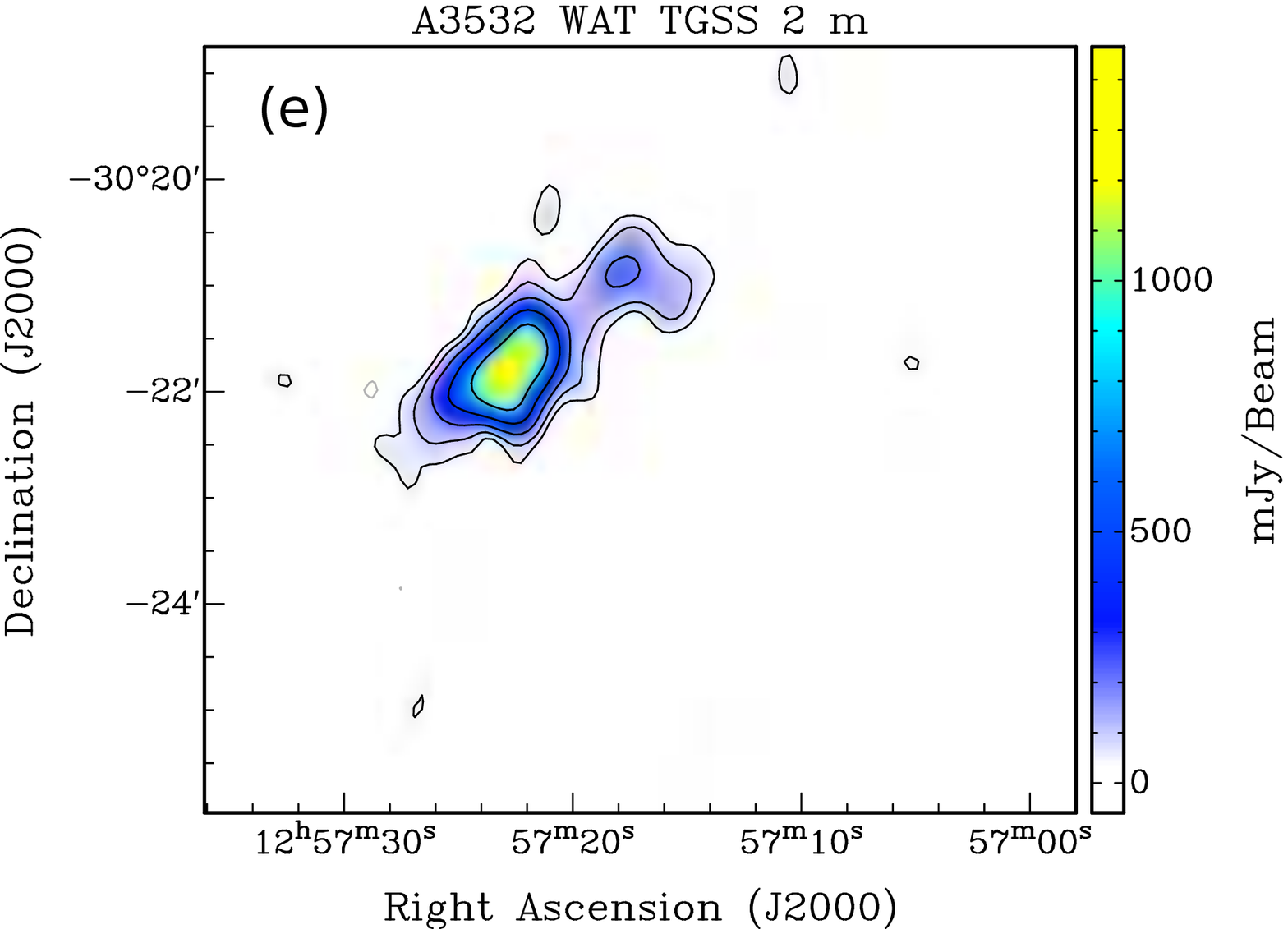}
\label{fig:A3532_TGSS_2m_WAT_source}
}
\subfigure
{
\includegraphics[width=2.15in]{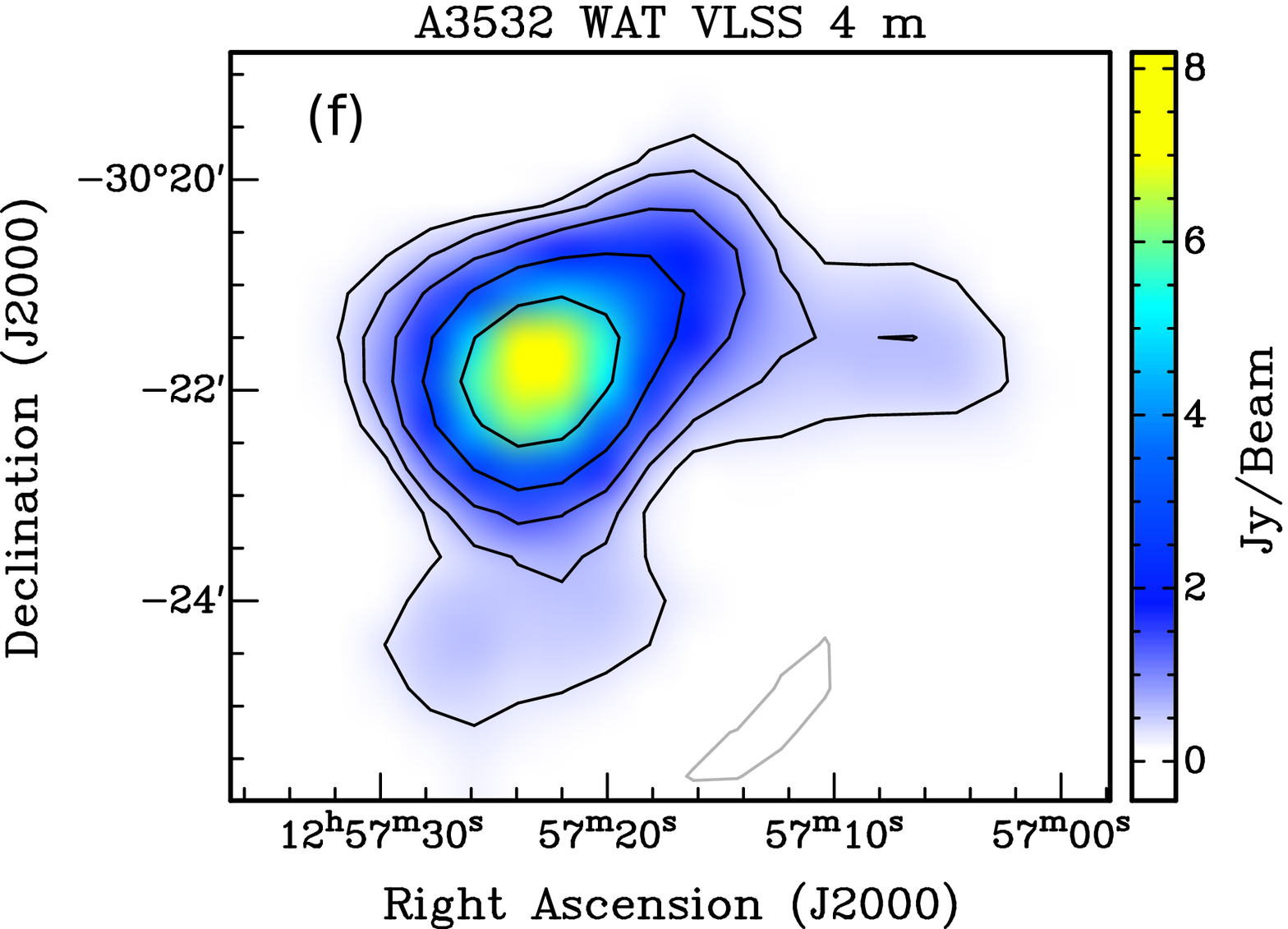}
\label{fig:A3532_VLSS_4m_WAT_source}
}\\
\subfigure
{
\includegraphics[width=1.8in]{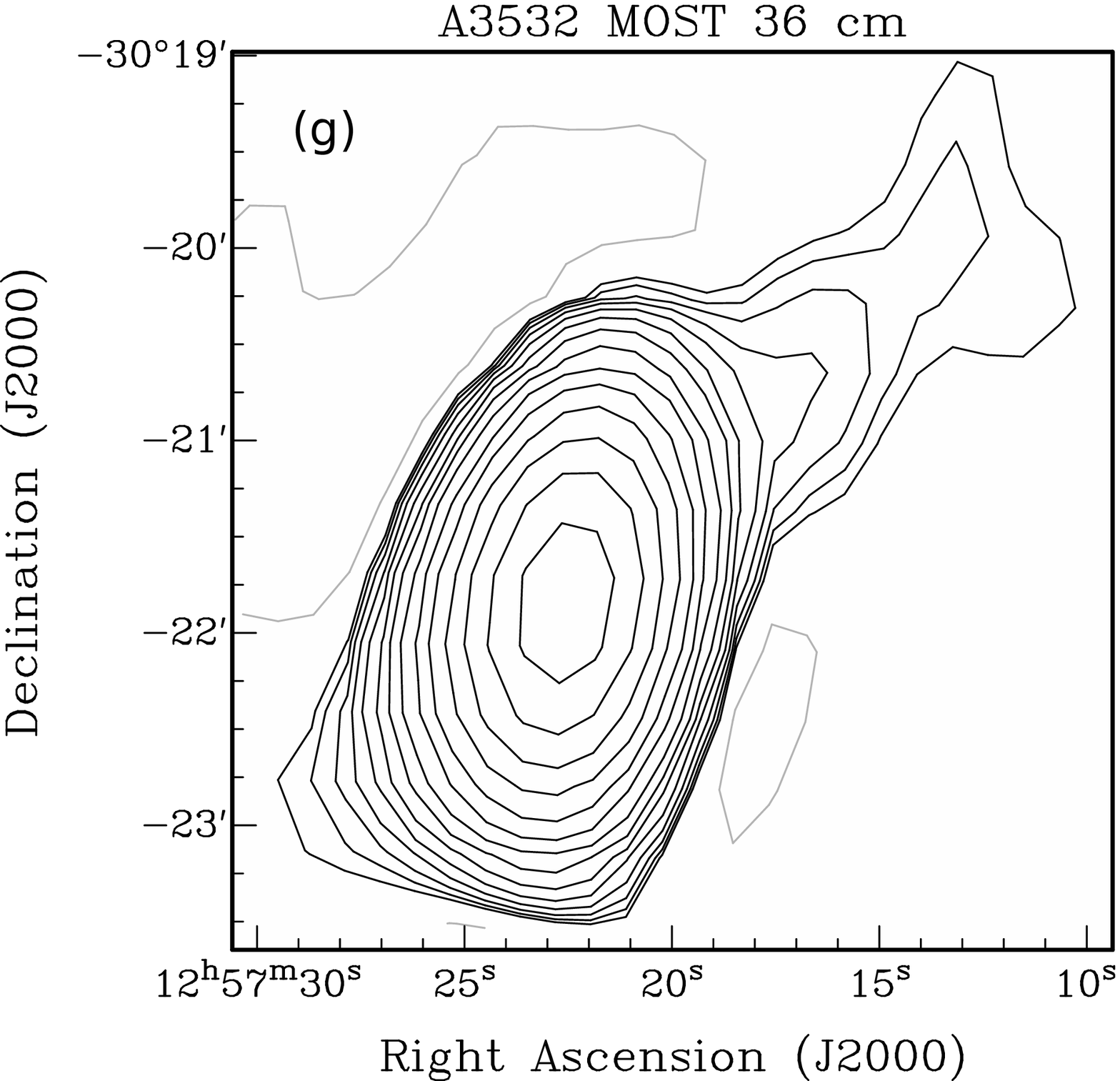}
\label{fig:A3532_MOST_36cm_WAT_source}
}
\caption{\small{The contoured radio images of the WAT source in A3532 from (a) VLA 20 cm, (b) ATCA 13 cm, (c) GMRT 50 cm, (d) VLA 6 cm , (e) TGSS 2 m 
and (f) VLSS 4m observations, with the contour levels starting at 0.3, 1, 1.5, 0.2, 50 and 300 mJy/beam, respectively, and increasing by 
factors of 2. (g): The contoured image of the WAT source from SUMSS 36 cm observation, with contour levels starting at 12 mJy/beam and increasing 
by factors of $\sqrt{2}$. The images have been arranged in increasing order of beam sizes. The hotspots \textquoteleft{}H1\textquoteright{} and
 \textquoteleft{}H2\textquoteright{}, and tails \textquoteleft{}T1\textquoteright{} and \textquoteleft{}T2\textquoteright{} of the WAT source, 
are marked in (a).}}
\label{fig:A3532_WAT_source_images}
\end{figure*}

   The overlay of the X-ray contours on the optical image of the two clusters from the SuperCOSMOS survey in the B$_{J}$ band is shown in 
Figure \ref{fig:A3532_A3530_SuperCOSMOS_optical_image}. It can be seen that the X-ray emission peaks of both clusters coincide with the 
positions of their respective BCGs. The 20 cm radio contours (red) in Figure \ref{fig:A3532_A3530_SuperCOSMOS_optical_image} show a 
\textquoteleft{}C\textquoteright{} shaped WAT source at the center of A3532, coincident with the position of the brighter nucleus of its 
dumbbell BCG. In Figure \ref{fig:A3532_WAT_source_images}, the WAT source is seen at all the radio wavelengths with sufficient angular 
resolution. The projected radio emission of the WAT galaxy is aligned in the NW-SE direction, i.e., approximately orthogonal to the major axis 
of the X-ray emission. The structure of the WAT is complicated somewhat by our viewing angle. It comprises two hotspots (locations of enhanced 
radio emission caused by the sudden slowing of the relativistic electrons as they emerge from the host galaxy into the intracluster medium) 
\textquoteleft{}H1\textquoteright{} and \textquoteleft{}H2\textquoteright{}, and two closely aligned tails \textquoteleft{}T1\textquoteright{} 
and \textquoteleft{}T2\textquoteright{} (Figure \ref{fig:A3532_VLA_20cm_WAT_source}), trailing back from the hotspots. The GMRT 50 cm, TGSS 2 
m and SUMSS 36 cm images (Figures \ref{fig:A3532_GMRT_50cm_WAT_source}, \ref{fig:A3532_TGSS_2m_WAT_source} and 
\ref{fig:A3532_MOST_36cm_WAT_source}, respectively) show an extension of the radio emission in the north-west direction. The VLSS observation 
also shows some emission in this part of the WAT but it is not well resolved.

 In Figures \ref{fig:A3532_VLA_20cm_WAT_source}, \ref{fig:A3532_atca_13cm_WAT_source} and \ref{fig:A3532_GMRT_50cm_WAT_source}, the tails of 
the WAT source appear to be very closely aligned and seem to eventually merge with each other. This makes the apparent morphology of 
the source intermediate to that of a WAT and a Narrow Angled Tailed (NAT) source. However, the NAT sources are usually associated with galaxies
 found at the cluster peripheries with their tails highly bent as a result of very high velocity ($\gtrsim$a few thousand km s$^{-1}$) motions 
of the host galaxies through the relatively stationary intracluster medium (ICM) (Jones \& Owen 1979; Klamer et al. 2004; Mao et al. 2009). On 
the other hand, the dominant cluster galaxies with which the WAT sources (like the one in A3532) are associated do not have peculiar velocities
 more than a few hundred km s$^{-1}$ and, therefore, the ram pressure due to the galaxy's motion is not very high (Burns 1981; Eilek et al. 
1984; O'Donoghue et al. 1990). It is also quite unlikely that the ram pressure from the ICM (resulting from the motions induced by the 
galaxy scale mergers), which  is generally responsible for the bent tails of the WAT sources (Eilek et al. 1984; Burns et al. 1994; Roettiger 
et al. 1996), could lead to the observed alignment of the WAT tails in A3532. Considering these facts, a likely interpretation is that the 
plane of the WAT is aligned at a small angle to our line of sight and the observed close alignment of the tails is due to the projection 
effects.

\begin{figure*}
\centering
\includegraphics[width=4in]{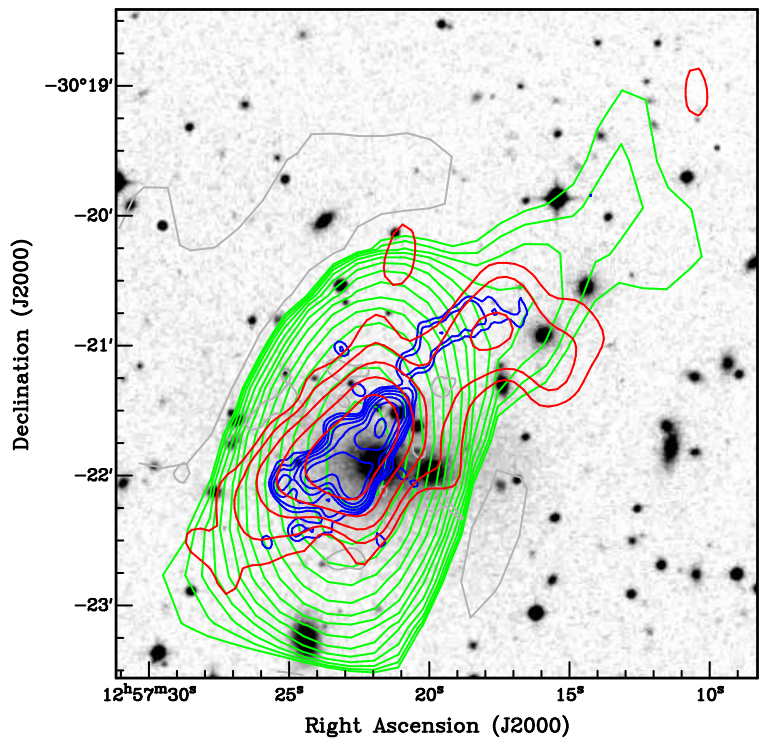}
\caption{The GMRT 2 m (TGSS; red) and 50 cm (blue), and SUMSS 36 cm (green) contours overlaid on the optical image of the central part 
of A3532 from the SuperCOSMOS survey. The north-western extensions of the WAT source seen in the three sets of radio contours seem to 
coincide.}
\label{fig:A3532_GMRT_50_cm_TGSS_MOST_843_MHz_ovld_optical}
\end{figure*}

 The north-western extension of the radio emission is only seen at the lowest frequencies. Figure 
\ref{fig:A3532_GMRT_50_cm_TGSS_MOST_843_MHz_ovld_optical} shows the GMRT 50 cm, TGSS 2 m, and SUMSS 36 cm contours overlaid on the optical 
image of A3532. The north-western extensions seen in the three sets of contours seem to coincide, strengthening the reality of the feature. 
However, the exact shapes and extents of these extensions do not match, mostly due to the different beam sizes. It should also be noted that in
 the TGSS survey observation, each source is observed for a very short duration of time ($\sim$3.5 minutes) and therefore, the final image may 
have certain artifacts. The extension does not seem to have an optical counterpart, therefore, it may either be a part of the WAT radio 
emission or a diffuse source unrelated to the WAT e.g, a radio relic in the cluster or a background source. Using the flux density estimates 
obtained from the GMRT 50 cm, TGSS and SUMSS observations, the extension is found to have a steep spectrum with a powerlaw spectral index, 
$\alpha\sim-2$. Deeper and high-resolution radio observations at lower frequencies will be required to make a detailed study of the morphology 
and spectral properties of the extension.

\begin{figure*}
\centering
\includegraphics[width=4.0in,angle=90]{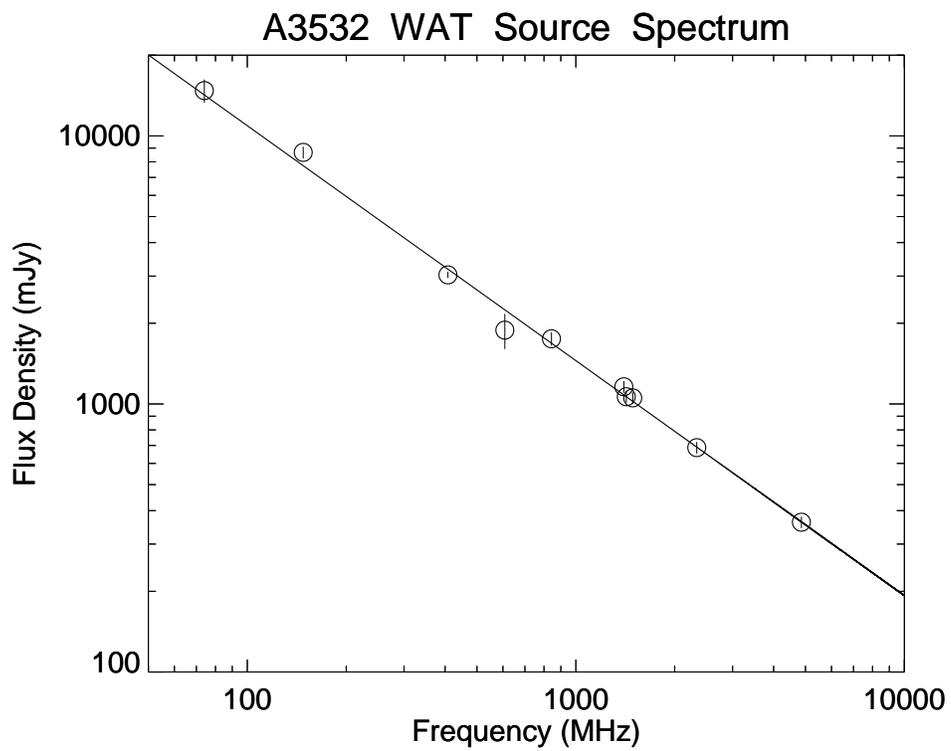}
\caption{The radio spectrum of the complete WAT source in A3532 on a log-log scale, fitted with a power law model. The flux densities used and 
the details of the observations are given in Table \ref{tab:radio_observation_table}.}
\label{fig:A3532_WAT_radio_spectrum}
\end{figure*}

 The estimated rms and flux densities of the full WAT source obtained from observations at various frequencies, are 
given in Table \ref{tab:radio_observation_table}. The spectrum of the WAT source (shown in Figure \ref{fig:A3532_WAT_radio_spectrum}) is 
consistent with a single powerlaw with a spectral index $\alpha$ of $-0.88\pm0.02$ (S$_{\nu}\propto \nu^{\alpha}$). The flux 
density of the WAT source obtained from the 50 cm GMRT observation seems to be a bit low, even after correcting for the primary beam pattern 
using the standard GMRT values. In this archival data set, the source was observed about 15$^{\prime}$ away from the phase center, and this 
along with the low declination of the source accounts for the large flux density error of $\sim$15\%. Also, the 20 cm flux densities obtained 
from the VLA and ATCA seem to be slightly low, compared with that from the NVSS, possibly a result of missing flux in high-resolution images.

\subsection{Global X-ray Spectra}
\label{sec:Total-Spec-Analy}
 We extracted \textit{XMM-Newton} spectra averaged over the cluster sizes for each of the clusters A3532 and A3530, and for the 
overlapping region (OR) between them. For A3532 and A3530, spectra were extracted from circular regions (radii $\sim9.6^{\prime}$ and 
$7.9^{\prime}$, respectively) centered on their respective X-ray emission peaks. For the overlapping region the spectra were extracted from a 
small elliptical region, common to both A3530 and A3532. For A3532, spectra from \textit{Chandra} were also 
extracted. As \textit{Chandra} did not cover the full circular region that was used to 
extract the \textit{XMM-Newton} spectra of A3532, a polygon shaped approximation to the circular region was made to extract the 
\textit{Chandra} spectrum. For this spectrum, an emission-free region near the cluster was used for extracting the background spectrum. 
The extraction of the background spectrum for the \textit{XMM-Newton} observations is described in \S\ref{sec:XMM_Background_treatment}.

\begin{figure*}
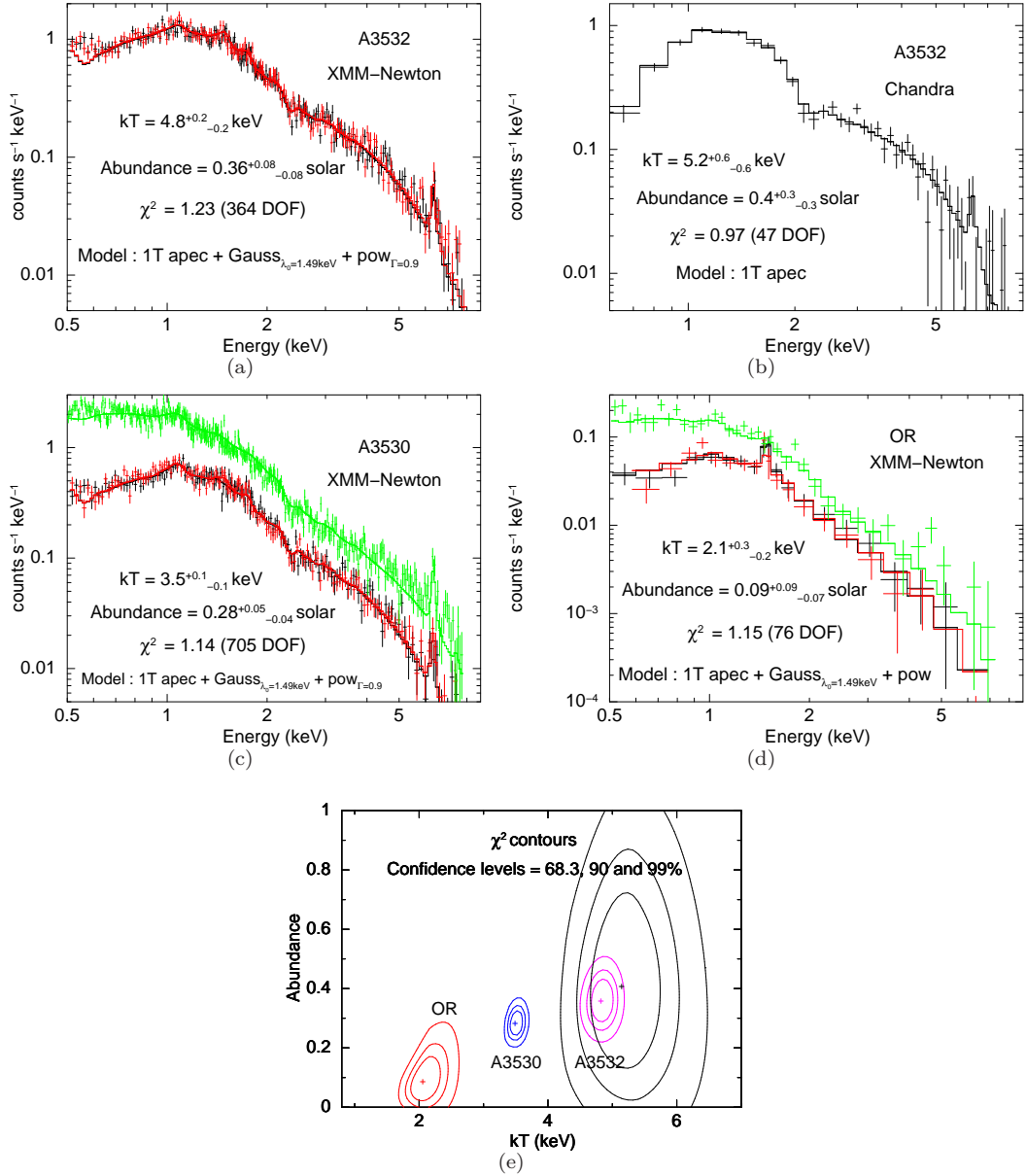

\centering
\subfigure[]
{
\includegraphics[width=1.9in,angle=270]{f7a.eps}
\label{fig:A3532_XMM_spectra}
}
\subfigure[]
{
\includegraphics[width=1.9in,angle=270]{f7b.eps}
\label{fig:A3532_Chandra_spectra}
}
\subfigure[]
{
\includegraphics[width=1.9in,angle=270]{f7c.eps}
\label{fig:A3530_XMM_spectra}
}
\subfigure[]
{
\includegraphics[width=1.9in,angle=270]{f7d.eps}
\label{fig:Filament_XMM_spectra}
}

\subfigure[]
{
\includegraphics[width=1.9in,angle=270]{f7e.eps}
\label{fig:A3532_A3530_filament_xmm_combined_chisq_cont}
}
\caption{\small{(a)-(c):Average spectra of the clusters A3532 and A3530, and their overlapping region (OR) from \textit{XMM-Newton} 
MOS1 (black), MOS2 (red) and PN (green) detectors. All the spectra have been fitted with \textit{wabs*apec} model shown as a histogram.
 Details of the spectral analysis are given in \S\ref{sec:Total-Spec-Analy}, and the best-fit parameters are shown here as 
insets. (d):Average spectra of A3532 from \textit{Chandra} data, fitted with the \textit{wabs*apec} model shown as a 
histogram. (e):The $\chi^{2}$ contours of the temperature and abundance measurements for the cluster A3532 (pink:\textit{XMM-Newton},
 black:\textit{Chandra}) and A3530 (blue:\textit{XMM-Newton}), and the overlapping region (OR) (red:\textit{XMM-Newton}). The confidence levels 
for the innermost, middle, and outermost contours for each of the 4 sets of contours are at 68.3\%, 90\% and 99\% respectively.}}
\label{fig:XMM_A3532_A3530_filament_average_spectra}
\end{figure*}

  The X-ray spectral fitting package \textit{XSPEC} (version 12.5.1) was used for all the spectral analyses. All spectra were fitted in the 
energy band 0.5-8.0 keV. The neutral hydrogen column densities along the line of sight to A3532 and A3530 were fixed to be 
$6.47\times 10^{20}$ cm$^{-2}$ and $6.24\times 10^{20}$ cm$^{-2}$, respectively, based on 
Leiden/Argentine/Bonn (LAB) Galactic HI survey (\cite{Kalberla:2005}) and the redshifts were frozen to their average values for the 
respective clusters. The \textit{wabs} photoelectric absorption model (\cite{MorrMcCam:1983}) and \textit{apec} plasma emission model 
(\cite{Smith:2001}) have been used for fitting all the spectra. The relative elemental abundances given in \cite{AndEbi:1982} were 
used for \textit{wabs}. For \textit{XMM-Newton} analyses, the MOS1, MOS2 and PN (only MOS1 and MOS2 for A3532) spectra were fitted 
simultaneously using three separate \textit{wabs*apec} models. For the spectra belonging to the same region, the values of abundance, 
temperature and \textit{apec} normalizations for the models were linked together but were kept free. To model the residual soft proton 
contamination, separate powerlaw models (see \cite{Snowden:2008}) were used, and to model the 
instrumental 1.49 keV Al K-$\alpha$ line, separate Gaussian components for MOS1, MOS2 and PN (see \cite{Snowden:2008}) were used. 
The \textit{Chandra} spectrum of 
A3532 was fitted using only the \textit{wabs*apec} model. The resulting spectra from all detectors, along with the histograms of the best-fit 
model spectra, are shown in Figures~\ref{fig:A3532_XMM_spectra}, \ref{fig:A3532_Chandra_spectra}, \ref{fig:A3530_XMM_spectra}, and 
\ref{fig:Filament_XMM_spectra}. The best-fit values of temperature, abundance and \textit{apec} normalizations are provided in Table 
\ref{tab:total_regions_spectral_results}.

 The confidence contours of the fitted temperatures and abundances, resulting from the spectral analyses at the 
68.3\%, 90\% and 99\% confidence levels are shown in Figure \ref{fig:A3532_A3530_filament_xmm_combined_chisq_cont}. It can be seen that the 
temperatures of the three regions are distinct at the 99\% confidence level. The cluster A3532 has the highest temperature ($4.8\pm0.2$ keV) 
and the overlapping region has the lowest temperature ($2.1^{+0.3}_{-0.2}$ keV). The abundance for the cluster A3532 is not distinct from that for 
the cluster A3530, even at the confidence level of 68.3\%. The overlapping region has a distinct and the the lowest abundance (at 90\% confidence 
level), among all the cluster regions. The \textit{Chandra} and \textit{XMM-Newton} results for A3532 are in good agreement, although results 
from \textit{Chandra} have larger errors.

\subsection{X-ray Luminosity estimates}
\label{sec:luminosity_estimates}
  X-ray luminosities of A3532 and A3530 in the energy range of 0.5-8.0 keV were estimated using the \textit{XMM-Newton} data 
(for A3532, using \textit{Chandra} data as well) from the flux values obtained from the spectral analysis of these regions described in 
\S\ref{sec:Total-Spec-Analy}. The fluxes ($\rm F_{\rm X}$) were estimated by convolving the model used in \S\ref{sec:Total-Spec-Analy} 
with the \textit{XSPEC} convolution model, \textit{cflux} after freezing the \textit{apec} normalization.
 The X-ray luminosities ($\rm L_{\rm X}$) were derived from the fluxes using the formula: 
\begin{equation} \label{eq:luminosity_equation}
\rm L_{\rm X} = 4 \pi \rm D_{\rm L}^{2} \rm F_{\rm X}
\end{equation}
where $\rm D_{\rm L}$ is the luminosity distance to the source. 
 The values of luminosities (L$_{\rm x}$) derived using this relation for the two clusters are given in Table 
\ref{tab:total_regions_spectral_results}. The 0.1-2.4 keV luminosities have also been estimated for the two clusters for comparison with 
values obtained by \cite{Ikebe:2002} by fitting 2-T thermal plasma models to the ASCA spectra of the clusters. The luminosities obtained by 
\cite{Ikebe:2002} have been scaled for the currently used value of the Hubble constant, i.e., 70 km s$^{-1}$ Mpc$^{-1}$. For the 0.1-2.4 keV 
luminosity estimates obtained by us, the values of the hydrogen column density and the redshifts have been frozen to the values used by Ikebe 
et al. for consistency. However, 2-T thermal plasma models could not be fitted to our spectra as the normalizations of the second apec 
components were negligibly small. We have, therefore, used only the 1-T apec models throughout our analyses. The 0.1-2.4 keV luminosities 
obtained by us are only slightly lower than those obtained by \cite{Ikebe:2002}, possibly due to the different spectral models used for the 
two analyses. These results have also been listed in Table \ref{tab:total_regions_spectral_results}.

\begin{figure*}
\centering
\includegraphics[width=5.0in,angle=270]{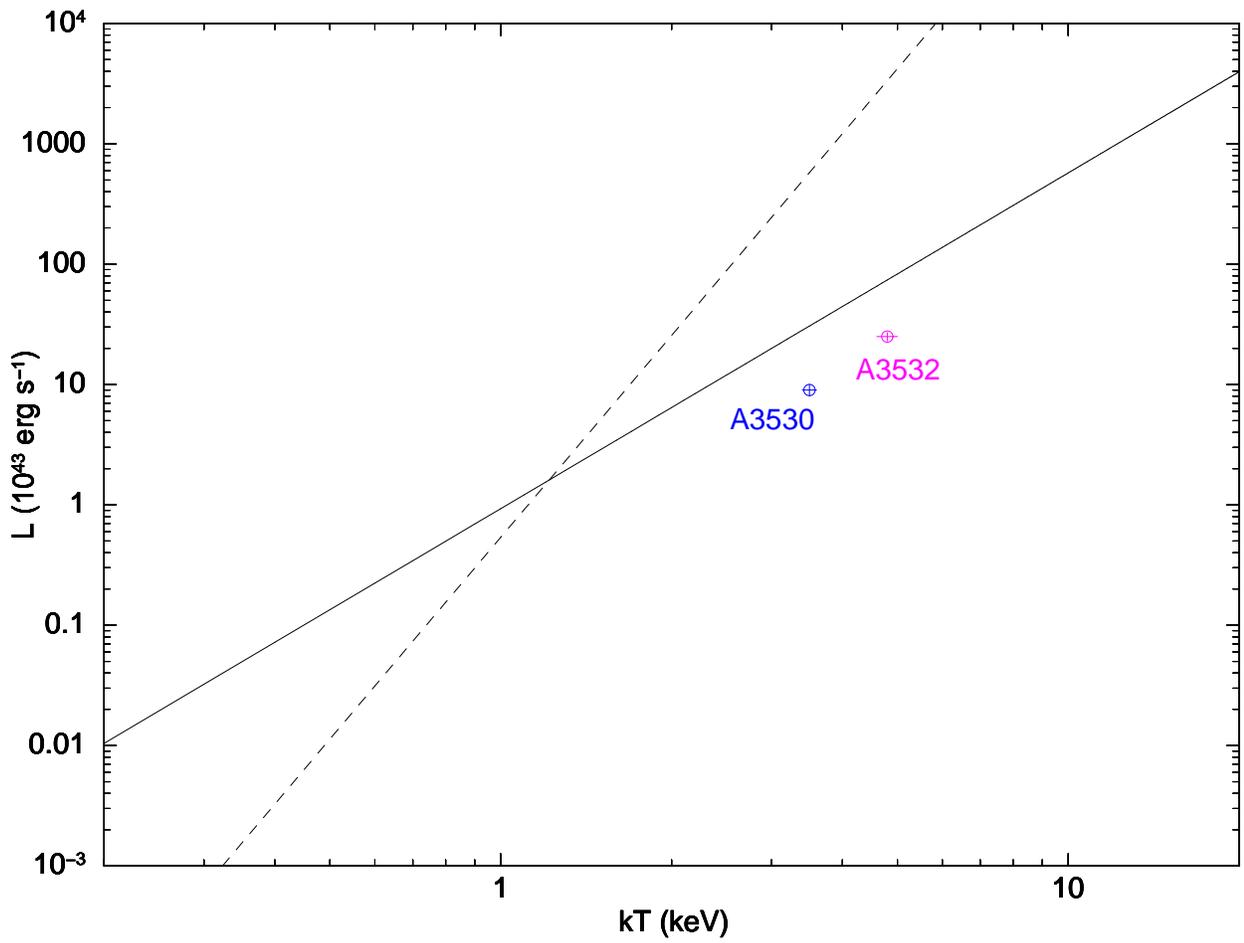}
\caption{The L$_{X}$-kT relations of rich clusters (solid line) and isolated groups of galaxies (dashed line), obtained by Xue \& Wu (2000). 
The positions of A3532 and A3530 have also been shown.}
\label{fig:A3532_A3530_Lx_kT_relation}
\end{figure*}

We have also estimated the bolometric X-ray luminosities of the clusters A3532 and A3530. The average count rates of the two clusters were 
obtained by fitting their $0^{\circ}-360^{\circ}$ surface brightness profiles with the $\beta$-model. These count rates were converted 
to fluxes in the 0.01-100 keV energy band, by using the HEASARC tool \textbf{ Web Portable, Interactive, Multi-Mission Simulator (WebPIMMS)}, 
from which the bolometric X-ray luminosities were estimated by using equation \ref{eq:luminosity_equation}. The results obtained from 
the $\beta$-model fitting along with the estimated X-ray bolometric luminosities of A3532 and A3530 have been given in Table 
\ref{tab:bol_xray_luminosity}. By analyzing the X-ray bolometric luminosities and temperatures of 274 clusters and 66 groups of galaxies, 
\cite{XuWu:2000} obtained their  L$_X$-kT relations as $\rm L_{X} = 10^{-0.032\pm0.065}T^{2.79\pm0.08}$ and 
$\rm L_{X} = 10^{-0.27\pm0.05}T^{5.57\pm1.79}$, for rich clusters and isolated groups of galaxies, respectively, where L$_{X}$ is in the units 
of 10$^{43}$ ergs s$^{-1}$, and T is in the units of keV. Figure \ref{fig:A3532_A3530_Lx_kT_relation} shows these relations using lines on a 
log-log plot along with the positions of A3532 and A3530. The two clusters lie very close to the line for the rich clusters of galaxies 
with only slightly higher temperatures.

\subsection{Radial Profiles of Thermodynamic Quantities based on Azimuthally Averaged Spectra}
\label{sec:Azimuth_spec_analys}
We have produced the azimuthally averaged profiles of temperature, density, entropy, and pressure by extracting spectra in eight circular 
annuli in each of the clusters A3532 and A3530 using \textit{XMM-Newton} data. For the cluster A3532, \textit{Chandra} data were also used for 
making these profiles. As \textit{Chandra} did not cover the full extent of the three outermost annuli in A3532, only the first five 
annuli were used for \textit{Chandra} data. The annuli centers were at the peak of the X-ray emission of each cluster, and the 
radius of the $n^{th}$ annulus was $n\times75^{\prime\prime}$ for both clusters. All spectra were extracted in the 0.5-8.0 keV energy band.
\subsubsection{Two-dimensional Projected Profiles}
\label{sec:projection_analysis} 
\begin{figure*}
\centering
\subfigure
{
\includegraphics[height=2.1in,angle=270]{f9a.eps}
\label{fig:A3532_proj_temp_anu_prof}
}
\subfigure
{
\includegraphics[height=2.1in,angle=270]{f9b.eps}
\label{fig:A3532_deproj_temp_anu_prof}
}
\subfigure
{
\includegraphics[height=2.3in,angle=270]{f9c.eps}
\label{fig:A3532_proj_dens_anu_prof}
}
\subfigure
{
\includegraphics[height=2.3in,angle=270]{f9d.eps}
\label{fig:A3532_deproj_dens_anu_prof}
}
\subfigure
{
\includegraphics[height=2.15in,angle=270]{f9e.eps}
\label{fig:A3532_proj_entr_anu_prof}
}
\subfigure
{
\includegraphics[height=2.15in,angle=270]{f9f.eps}
\label{fig:A3532_deproj_entr_anu_prof}
}
\subfigure
{
\includegraphics[height=2.25in,angle=270]{f9g.eps}
\label{fig:A3532_proj_press_anu_prof}
}
\subfigure
{
\includegraphics[height=2.25in,angle=270]{f9h.eps}
\label{fig:A3532_deproj_press_anu_prof}
}
\subfigure
{
\includegraphics[height=1.4in]{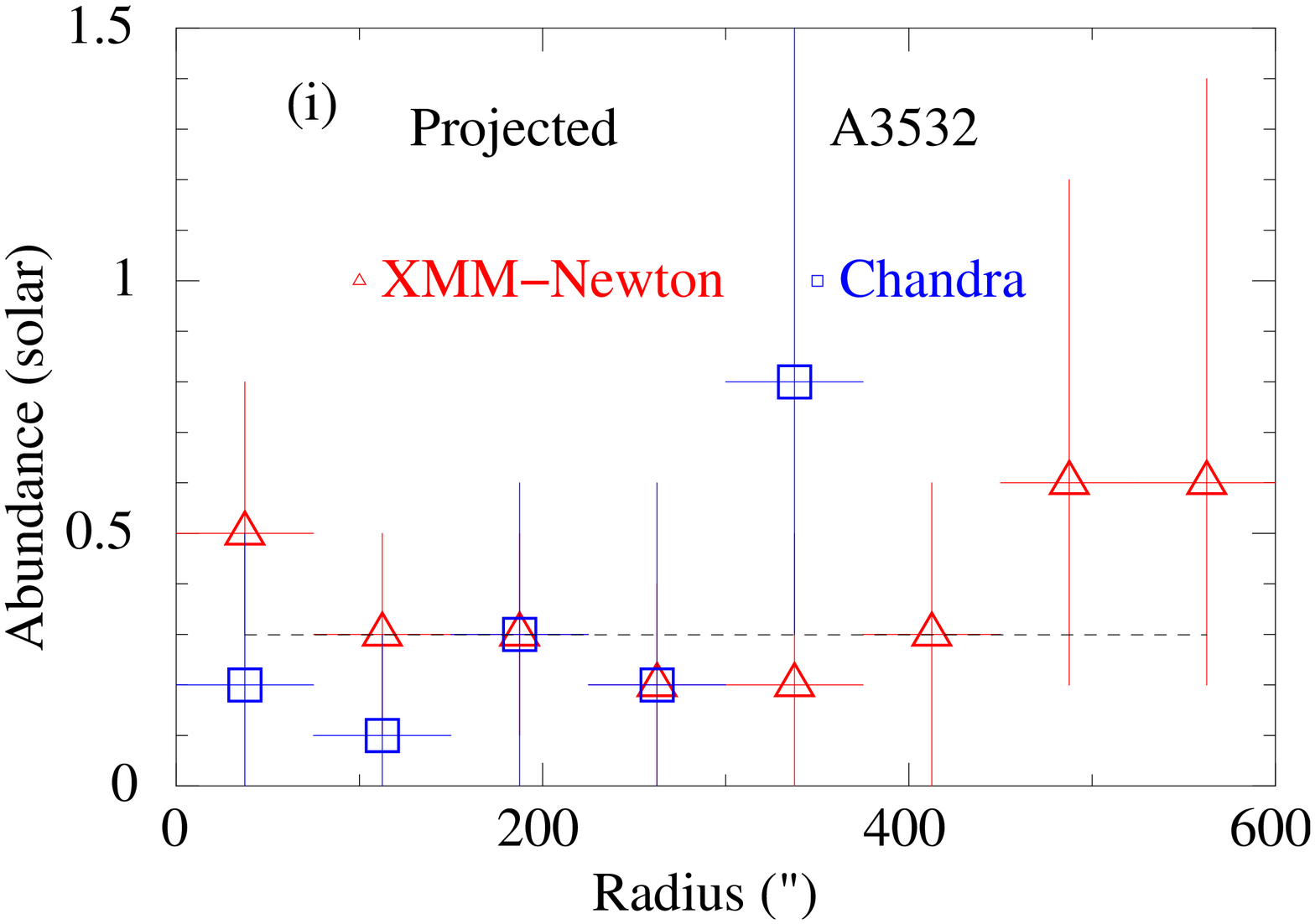}
\label{fig:A3532_proj_abund_anu_prof}
}
\caption{\small{(a)-(h): Projected and deprojected temperature (kT), electron density (n$_{e}$), entropy (S), and pressure (P)  profiles obtained 
from the spectral analysis of the \textit{XMM-Newton} MOS spectra from eight circular annuli (red points) and \textit{Chandra} spectra from 
five circular annuli (blue points) in the cluster A3532. The value of elemental abundance was frozen to 0.36 times the solar value. (i): The 
abundance profile of A3532 from projected spectral analysis, after freeing the abundance parameter. The details of the projected and 
deprojected spectral analysis are given in \S\ref{sec:projection_analysis} and \S\ref{sec:deprojection_analysis}, respectively.}}
\label{fig:A3532_proj_deproj_anu_prof}
\end{figure*}

\begin{figure*}
\centering
\subfigure
{
\includegraphics[height=2.05in,angle=270]{f10a.eps}
\label{fig:A3530_proj_temp_anu_prof}
}
\subfigure
{
\includegraphics[height=2.05in,angle=270]{f10b.eps}
\label{fig:A3530_deproj_temp_anu_prof}
}
\subfigure
{
\includegraphics[height=2.25in,angle=270]{f10c.eps}
\label{fig:A3530_proj_dens_anu_prof}
}
\subfigure
{
\includegraphics[height=2.25in,angle=270]{f10d.eps}
\label{fig:A3530_deproj_temp_anu_prof}
}
\subfigure
{
\includegraphics[height=2.15in,angle=270]{f10e.eps}
\label{fig:A3530_proj_entr_anu_prof}
}
\subfigure
{
\includegraphics[height=2.15in,angle=270]{f10f.eps}
\label{fig:A3530_deproj_temp_anu_prof}
}
\subfigure
{
\includegraphics[height=2.2in,angle=270]{f10g.eps}
\label{fig:A3530_proj_press_anu_prof}
}
\subfigure
{
\includegraphics[height=2.2in,angle=270]{f10h.eps}
\label{fig:A3530_deproj_press_anu_prof}
}
\subfigure
{
\includegraphics[height=1.4in]{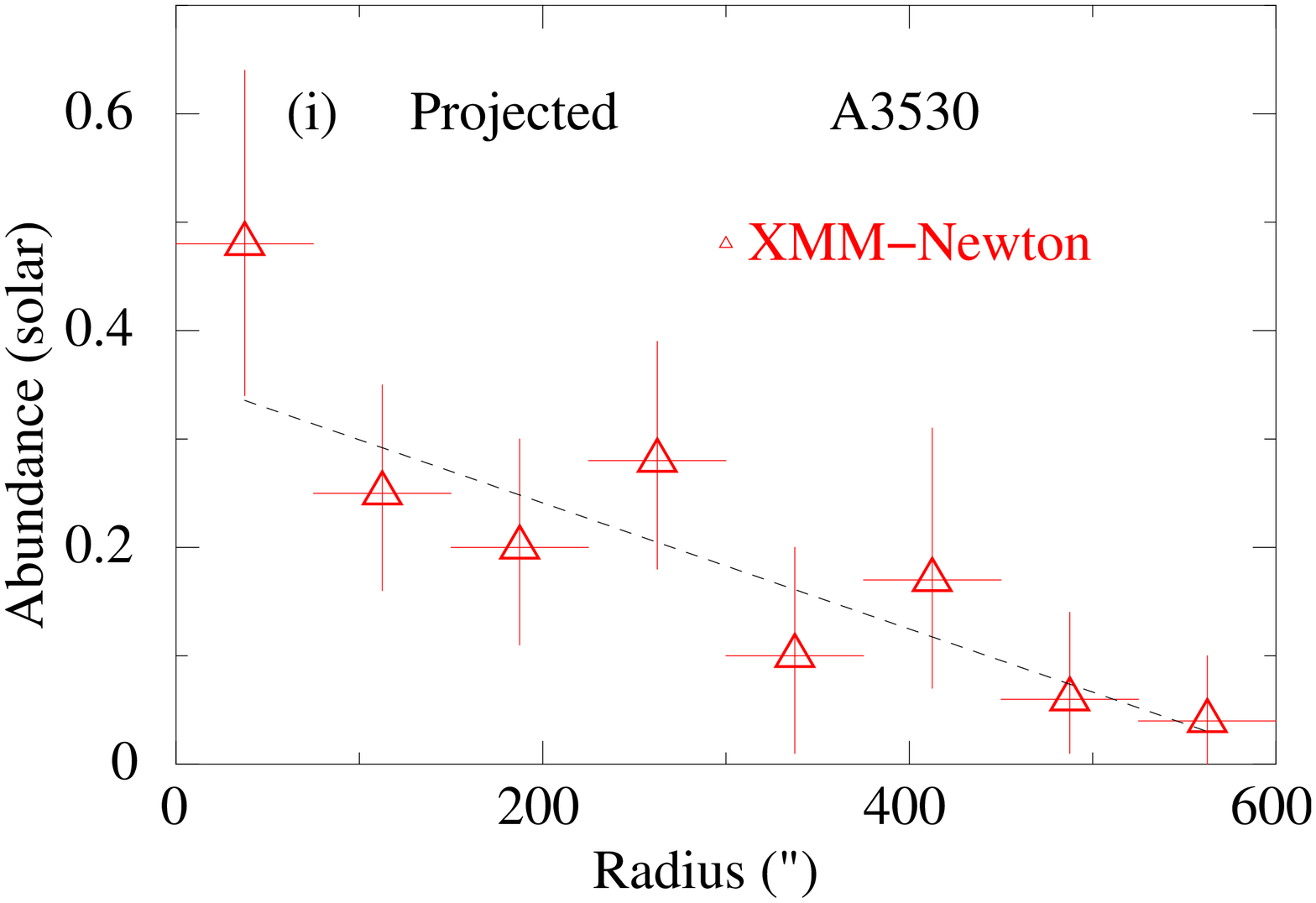}
\label{fig:A3530_proj_abund_anu_prof}
}
\caption{(a)-(h): Projected and deprojected temperature (kT), electron density (n$_{e}$), entropy (S), and pressure (P)  profiles obtained 
from the spectral analysis of the \textit{XMM-Newton} (MOS+PN) spectra from eight circular annuli in the cluster A3530. (i): The abundance 
profile of A3530 from projected spectral analysis. For the deprojected spectral analysis, the value of elemental abundance was frozen to 
0.28 times the solar value. The details of the projected and deprojected spectral analysis are given in 
\S\ref{sec:projection_analysis} and \S\ref{sec:deprojection_analysis}, respectively.}
\label{fig:A3530_proj_deproj_anu_prof}
\end{figure*}

  To determine the 2-D profiles, we used the same method as given in \S\ref{sec:Total-Spec-Analy} for spectral analysis. As no significant 
variations were seen in the elemental abundances of the annuli belonging to A3532, their values were frozen to that obtained in 
\S\ref{sec:Total-Spec-Analy}, i.e., 0.36 times the solar value ($Z_{\odot}$) (Table \ref{tab:total_regions_spectral_results}). In A3530, 
the elemental abundance was found to be significantly varying, therefore, for the spectral analysis of the annuli belonging to A3530, the 
elemental abundance was kept as a free parameter. We have verified for both clusters that while results do not change significantly for a free 
or frozen elemental abundance, their errors decrease slightly for a frozen abundance. For \textit{XMM-Newton} 
spectra, the residual soft proton contamination and instrumental Al lines were modeled by adding powerlaws and Gaussian components to the 
models as was also done in \S\ref{sec:Total-Spec-Analy}. Projected profiles of temperature (kT), density (n$_{e}$), entropy (S), and pressure 
(P) were produced and are shown as the left hand side images in Figures \ref{fig:A3532_proj_deproj_anu_prof} and 
\ref{fig:A3530_proj_deproj_anu_prof}. The temperature profiles were obtained as a direct result of the
 spectral analyses and are shown in Figures \ref{fig:A3532_proj_temp_anu_prof} and \ref{fig:A3530_proj_temp_anu_prof} for A3532 and A3530, 
respectively. To derive the electron density n$_{e}$, we used the \textit{apec} normalizations, 
$K=10^{-14}EI/(4\pi[D_{A}(1+z)]^{2}$), where 
$EI$ is the emission integral $\int n_{e} n_{p}dV$. By assuming, $n_{p}= 0.855 n_{e}$ (\cite{Henry:2004}) and a constant density within each 
spherical shell, we obtain, $EI=0.855n_{e}^{2}$V, where $V$ is the volume of spherical shell ($=4\pi(r_{o}^{3}-r_{i}^{3}$), where $r_{o}$ and 
$r_{i}$ are the radii of the inner and outer annuli forming the shell). The resulting density profiles are shown in Figures 
\ref{fig:A3532_proj_dens_anu_prof} and \ref{fig:A3530_proj_dens_anu_prof}, for A3532 and A3530, respectively. The entropy ($S$) and the 
electron pressure ($P$) are obtained from the relations, $S=kTn_{e}^{-2/3}$ and $P=n_{e}kT$, respectively (\cite{Gitti:2010}). The resulting 
entropy profiles for A3532 and A3530 are shown in Figures \ref{fig:A3532_proj_entr_anu_prof} and \ref{fig:A3530_proj_entr_anu_prof}, and the 
pressure profiles are shown in Figure \ref{fig:A3532_proj_press_anu_prof} and \ref{fig:A3530_proj_press_anu_prof}, respectively. The values of 
all the thermodynamic quantities for each of the annuli are tabulated in Tables \ref{tab:A3532_projected_annuli_spectral_results} and 
\ref{tab:A3530_projected_annuli_spectral_results} for A3532 and A3530, respectively. The temperature profiles obtained using 
\textit{XMM-Newton} data show low temperatures in the innermost annuli of both the clusters. For rest of the annuli in A3532, temperature 
is almost a constant, whereas for A3530, the temperature profile shows a gradual decrease outwards. The density, entropy and pressure profiles 
of both the clusters show an average decrease, increase and decrease, respectively, from the innermost to the outermost annulus. The projected 
profile of the elemental abundance in A3532 is shown in Figure \ref{fig:A3532_proj_abund_anu_prof}, and due to large errors, does not show 
significant variations in the values obtained for different annuli. The elemental abundance in A3530 shows a gradual decrease from the 
innermost to the outermost annulus (Figure \ref{fig:A3530_proj_abund_anu_prof}), which indicates an enrichment of the ICM towards the center of
 the cluster. The errors in the abundance profile of A3532 obtained using XMM-Newton spectra are much larger than those for A3530. This is 
because only MOS spectra were used for A3532 while both MOS and PN spectra were used for A3530. All the profiles of thermodynamic quantities 
obtained using \textit{Chandra} data are consistent with those obtained using \textit{XMM-Newton} data.

\subsubsection{Deprojected Profiles}
\label{sec:deprojection_analysis} 
  To get a better idea of the variations in the thermodynamic quantities which may get smoothed out due to projection effects , we carried out 
a deprojection analysis on the annuli described in \S\ref{sec:projection_analysis}. For this purpose, we used the \textit{XSPEC projct} model, 
which can estimate the parameters in 3-D space from the two-dimensional (2D) projected spectra of annular ellipsoidal shells, along with the 
\textit{wabs*apec} model. As the elemental abundances for all the annuli belonging to A3532 and A3530 did not show significant variations, 
their values were frozen to 0.36 and 0.28 times the solar value ($Z_{\odot}$), respectively. For \textit{XMM-Newton} spectra, the residual 
soft proton contamination and instrumental Al lines were modeled by adding powerlaws and Gaussian components to the models as was also done in 
\S\ref{sec:Total-Spec-Analy}. Note that, as the \textit{projct} model requires all the spectra belonging to the same annulus to be part of the 
same group, for each annulus a single powerlaw and Gaussian were used for all MOS1, MOS2 and PN detectors. Electron density 
(n$_{e}$), entropy (S), and electron pressure (P) have been calculated using the same relations as given in \S\ref{sec:projection_analysis}. 
The resulting deprojected profiles of temperature, density, entropy and pressure obtained for A3532 and A3530, are shown plotted on the right 
hand sides of Figures \ref{fig:A3532_proj_deproj_anu_prof} and \ref{fig:A3530_proj_deproj_anu_prof}, and listed in Tables 
\ref{tab:A3532_deprojected_annuli_spectral_results} and \ref{tab:A3530_deprojected_annuli_spectral_results}, respectively.

  The deprojected profiles of both clusters have errors larger than the projected profiles. The \textit{XMM-Newton} temperature profiles of
 the clusters A3532 and A3530 do not change significantly and are almost constant from the innermost to the outermost annulus, except for 
the innermost annulus of A3530, which shows a significantly lower temperature. On average, the density, entropy and pressure profiles of both 
the clusters show a gradual decrease, increase and decrease, respectively, from the innermost to the outermost annulus, as in their projected 
profiles. Results from \textit{Chandra} and \textit{XMM-Newton} for the cluster A3532 are in good agreement with each other. However, the 
errors from \textit{Chandra} data are much larger than those from \textit{XMM-Newton}, as also observed for the projected profiles. The 
projected and deprojected profiles of both the clusters do not seem to be significantly different, except for a few anomalies. The density 
values in the inner annuli of both the clusters, obtained from the projected spectral analysis, are higher than those from the deprojected 
spectral analysis. Similarly, the pressure values in the inner annuli of both the clusters, obtained from the XMM-Newton projected spectral 
analysis, are higher than those from the deprojected spectral analysis.

\subsection{Spectrally Determined 2D Projected Thermodynamic Maps at a Higher Resolution}
\label{sec:box_thermodynamic_maps}
 The 2D projected temperature, density, entropy, and pressure maps for the combined system of the A3532 and A3530 clusters, 
have also been made, using \textit{XMM-Newton} spectra from a total of 77 box shaped regions. As \textit{Chandra} data (available only for 
the A3532 cluster) had large errors (evident in \S\ref{sec:Total-Spec-Analy}, \S\ref{sec:projection_analysis} and 
\S\ref{sec:deprojection_analysis}), we have not used the \textit{Chandra} data in this Section. Out of the 77 box regions, 41 boxes were from 
A3532 and 37 were from A3530, with one box in common. An adaptive approach was followed for choosing the sizes of the boxes, so
 as to get sufficient counts in each region. Large size boxes ($\sim 7.7^{\prime}\times3.8^{\prime}$) for the outermost parts, small size boxes
 ($\sim1.9^{\prime}\times1.9^{\prime}$) for the innermost brightest parts, and medium sized boxes ($\sim 1.9^{\prime}\times3.8^{\prime}$) for 
the regions in between were selected in order to get more than 700 total counts from all three detectors in each box. Spectra from all  
boxes were fitted using \textit{wabs*apec} model with fixed Galactic absorption. As described in \S\ref{sec:Total-Spec-Analy}, the 
residual soft proton contamination and instrumental Al lines were modeled by adding powerlaws and Gaussian components to the models. Because 
of the poor statistics and large errors in the abundance values, the elemental abundances for all the box regions were fixed to the 
average abundance value of the cluster (see \S\ref{sec:Total-Spec-Analy}) to which the box belonged i.e., 0.36 and 0.28 times the 
solar value ($Z_{\odot}$) for the boxes belonging to A3532 and A3530, respectively. The electron density, entropy, and electron pressure 
were calculated using the same relations as in \S\ref{sec:projection_analysis}. Spherical geometry was assumed for the volume 
calculation. All 77 box regions were assumed to be projections of parts of spherical shells (centered at the X-ray intensity peak of 
the cluster to which the box belongs) with inner and outer radii ($R_{in}$, $R_{out}$) equal to the smallest and largest distance from the 
center of their respective spheres. The volume for each box region was estimated as 
$D_{A}^{3} \Omega (\theta_{out}^{2} - \theta_{in}^{2})^{1/2}$ (\cite{Henry:2004}; \cite{Ehlert:2011}), where $D_{A}$ 
is the angular diameter distance of the cluster to which the box belongs and $\Omega$ is the solid angle subtended by the box. 
$\theta_{in}$ and $\theta_{out}$ are equal to the distances R$_{in}$ and R$_{out}$ expressed in angular units respectively. For a box region 
common to both A3532 and A3530 an average of results from the two observations was used.

\begin{figure*}
\centering
\subfigure[]
{
\includegraphics[width=5.8in]{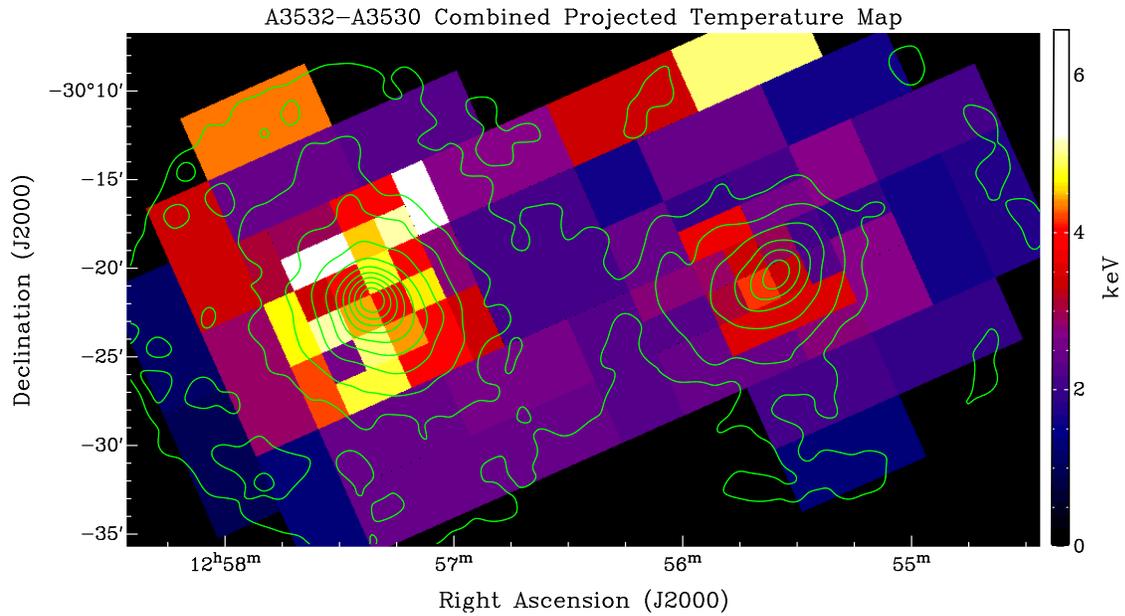}
\label{fig:2d_temp_map}
}
\subfigure[]
{
\includegraphics[width=5.8in]{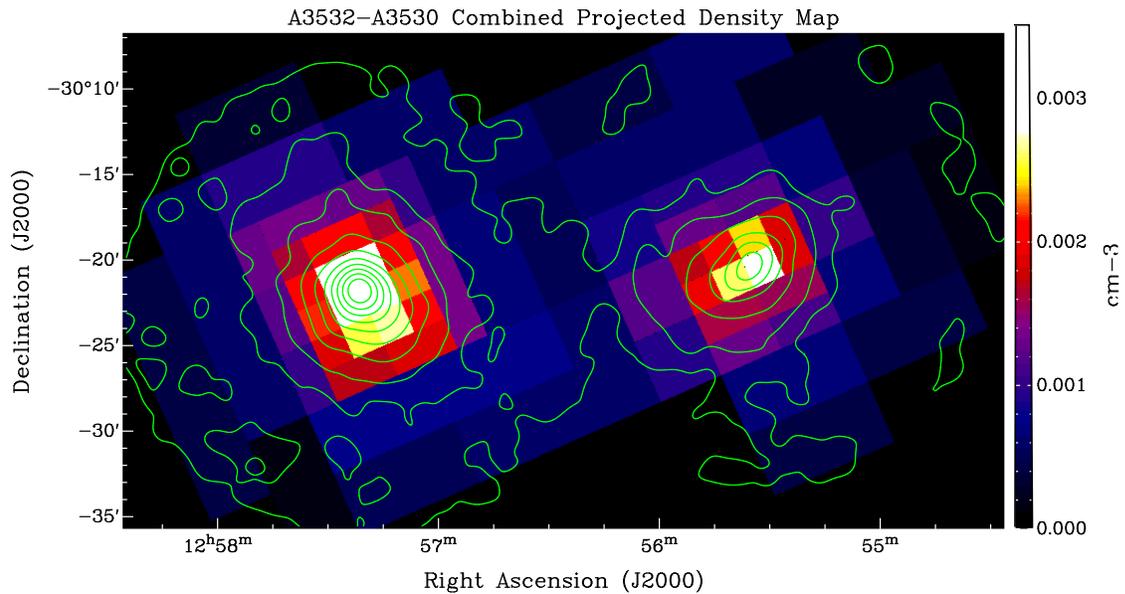}
\label{fig:2d_density_map}
}
\caption{(a): Projected temperature (kT) map from 77 box regions in the clusters A3532 and A3530 using \textit{XMM-Newton} data. The scale is 
expressed in keV units and is shown in the bar alongside. (b): Projected density (n$_{e}$) map from 77 box regions in the clusters A3532 and 
A3530 using \textit{XMM-Newton} data. The scale is expressed in the units of cm$^{-3}$ and is shown in the bar alongside. X-ray 
surface-brightness contours, with levels same as in Figure~\ref{fig:A3532_A3530_combined_MOS_image}, have been overlaid on both the figures. 
Details of the spectral fittings are provided in Section \ref{sec:box_thermodynamic_maps}.}
\label{fig:XMM_A3532_A3530_temp_density_maps}
\end{figure*}

\begin{figure*}
\centering
\subfigure[]
{
\includegraphics[width=5.8in]{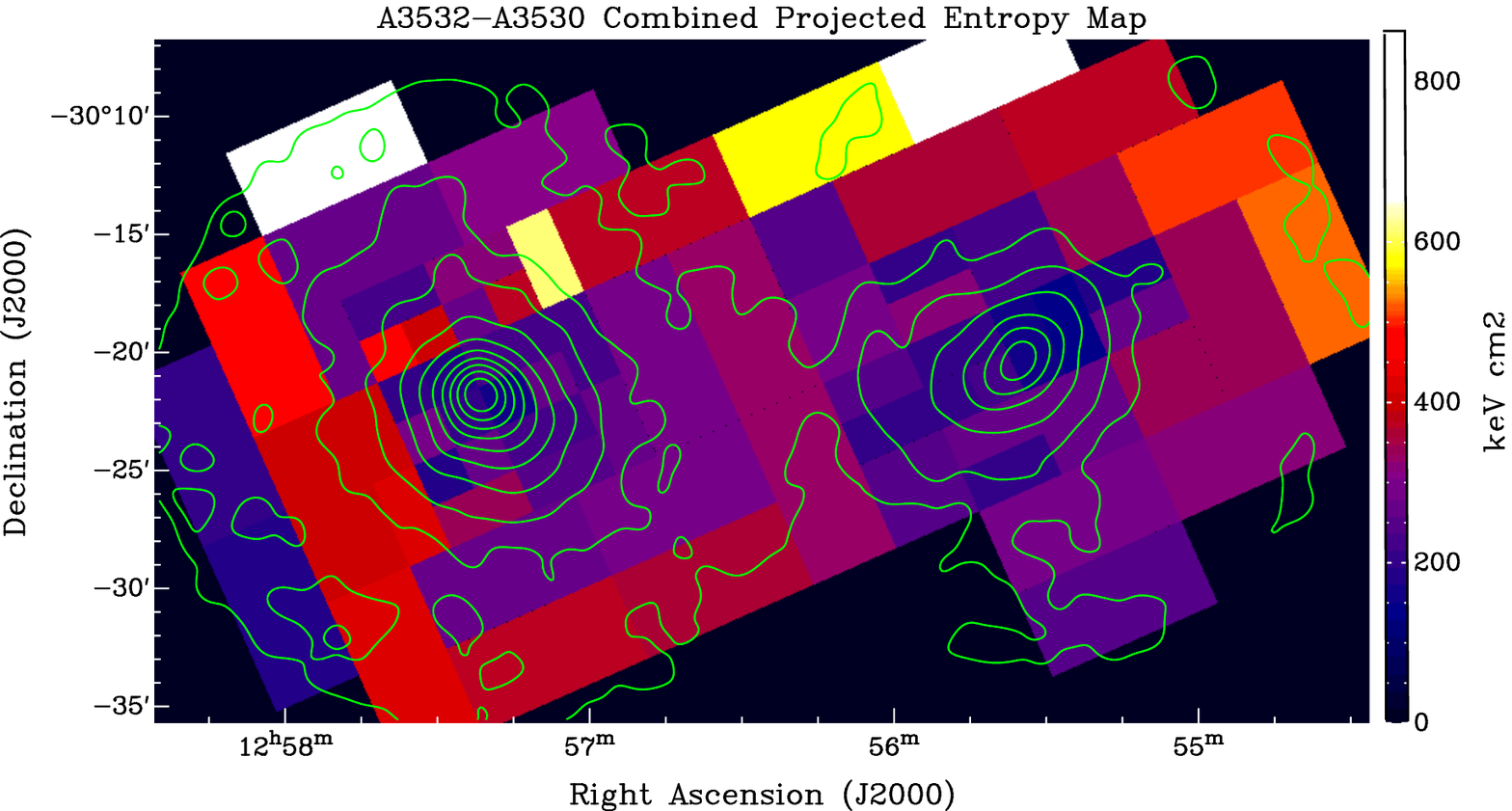}
\label{fig:2d_entropy_map}
}
\subfigure[]
{
\includegraphics[width=5.8in]{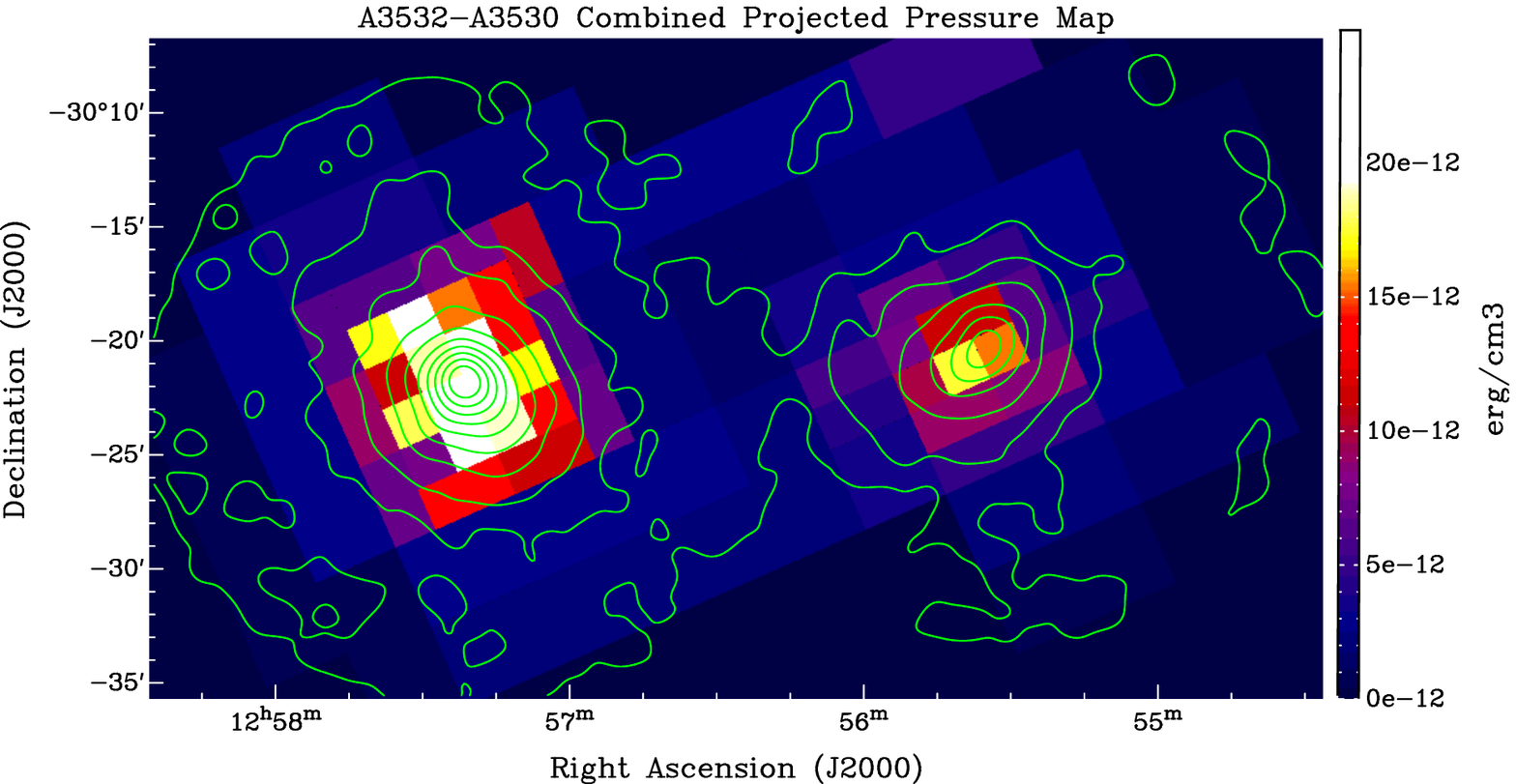}
\label{fig:2d_pressure_map}
}
\caption{(a): Projected entropy map from 77 box regions in the clusters A3532 and A3530 using \textit{XMM-Newton} data. The scale is expressed 
in units of keV cm$^{2}$ and is shown in the bar alongside. (b): Projected pressure map from 77 box regions in the clusters A3532 and A3530 
using \textit{XMM-Newton} data. The scale is expressed in the units of erg cm$^{-3}$ and is shown in the bar alongside. X-ray 
surface-brightness contours, with levels same as in Figure~\ref{fig:A3532_A3530_combined_MOS_image}, have been overlaid on both the figures. 
Details of the spectral fittings are provided in Section \ref{sec:box_thermodynamic_maps}.}
\label{fig:XMM_A3532_A3530_entropy_pressure_maps}
\end{figure*}

  The temperature, density, entropy, and pressure maps produced are shown in Figures~\ref{fig:2d_temp_map}, \ref{fig:2d_density_map}, 
\ref{fig:2d_entropy_map} and \ref{fig:2d_pressure_map}, respectively. The temperature in both clusters appears to decrease as we move outwards 
from the center. However, A3532 shows a lot of anisotropic variations in the temperature, especially in its central parts, though the 
statistical significance is low. Both density and pressure maps show a peak at the center of the clusters followed by an almost uniform 
decrease outwards. The entropy maps of both clusters show the presence of a few high entropy regions in their outer parts while almost a 
constant entropy is observed in their inner parts. The overlapping region between the two clusters does not show the presence of high 
temperature or high entropy, as would have been expected if an active merger was taking place between the two clusters. We have also made an 
estimate of the density ($n_{e}$ in the overlapping region by using the apec normalization obtained in \S\ref{sec:Total-Spec-Analy}. For volume 
calculation we assumed a prolate ellipsoid made using the ellipse used in \S\ref{sec:Total-Spec-Analy} for the overlapping region. We obtained 
a density of $(6.4\pm0.4)\times10^{-4}$ cm$^{-3}$ for the overlapping region, which is consistent with the its value from Figure 
\ref{fig:2d_density_map}.

\subsection{Cooling time}
\label{sec:cool_time_estimate}
     A commonly used relation for estimating the cooling time of a cluster from \cite{Sarazin:1988}, is as follows:
\begin{equation}
\label{eqn:cooling_time}
\rm t_{\rm cool}= 8.5 \times 10^{10} \rm yr \left[ \frac{\rm n}{10^{-3} \rm cm^{-3}}\right]^{-1} \left[\frac{\rm T_{\rm g}}{10^{8}K} \right]^{1/2}.
\end{equation}
 Using this relation and the central gas temperatures ($\rm T_{\rm g}$) and densities (n) (derived from the deprojection analysis in 
\S\ref{sec:deprojection_analysis}, the cooling times estimated for both A3532 and A3530 ($=1.8\times 10^{10}$y and $1.7\times 10^{10}$y, 
respectively) seem to be longer than the Hubble time ($\sim1.35\times 10^{10}$y). Note that, the innermost annuli in 
\S\ref{sec:deprojection_analysis} have radii equal to 75$^{\prime\prime}$ ($\sim$80 kpc). It should also be noted that the above relation is 
derived by assuming thermal bremsstrahlung as the only cooling mechanism. However, additional cooling by line emission may result in a smaller 
value of the cooling time. Using the continuum and line emissivity relations given in Sarazin (1988), the cooling times of A3530 
($\sim1.02\times 10^{10}$y) seems to be slightly lower while that of A3532 ($\sim1.26\times 10^{10}$y) seems to be very close to the Hubble 
time. A discussion regarding the possibility of cool cores in the cluster pair, based on the results 
obtained in this section along with some other results, is given in \S\ref{sec:discussion}.

\subsection{Gas Mass Estimation}
\label{sec:gas_mass_estimates}
 We have estimated the gas masses for A3532 and A3530 by using the gas density profiles obtained in \S\ref{sec:projection_analysis} and 
\S\ref{sec:deprojection_analysis}. The projected and deprojected gas density profiles for both clusters were fitted using a $\beta$-model 
i.e., 
\begin{equation}
\rm n_{\rm e}(\rm r)=\rm n_{\rm e}\left(0\right)\left(1+\frac{\rm r^{2}}{\rm r_{\rm c}^{2}}\right)^{(3/2)\rm \beta},
\end{equation}
 where $\rm n_{\rm e}(0)$ is the central density and $\rm r_{\rm c}$ is the 
core radius. The gas masses $\rm M_{\rm gas}(r)$ out to radii 0.5 Mpc and 1 Mpc for the 
two clusters were obtained by using the following formula (see \cite{Donnelly:2001}) : 
\begin{equation}
\rm M_{\rm gas}(\rm r)=4 \pi \rho_{0} \int_{0}^{\rm r} \rm s^{2} \left[1+ \left( \frac{\rm s}{\rm r_{\rm c}} \right)^{2} \right] ^{(3/2)\rm \beta} ds
\end{equation} 
where $\rho_{0}= \rm \mu \rm n_{\rm e}(0) \rm m_{\rm p}$, $\rm m_{\rm p}$ is the mass of a proton, and $\mu$= 0.609 is the average 
molecular weight for a fully ionized gas (\cite{Gu:2010}). The values of $\rm \beta$, $\rm r_{\rm c}$, 
$\rm \rho_{0}$, and $\rm M_{\rm gas }$ based on fitting the density profiles with the above model are listed in 
Table~\ref{tab:gas_mass_estimate}. the results obtained from both projected and deprojected analysis show A3532 to be marginally more massive 
than A3530.

\subsection{Galaxy Velocity Distribution and Virial Mass}
\label{sec:gal_vel_distrb_an_vir_mas}
\begin{figure*}
\centering
\includegraphics[width=5.0in]{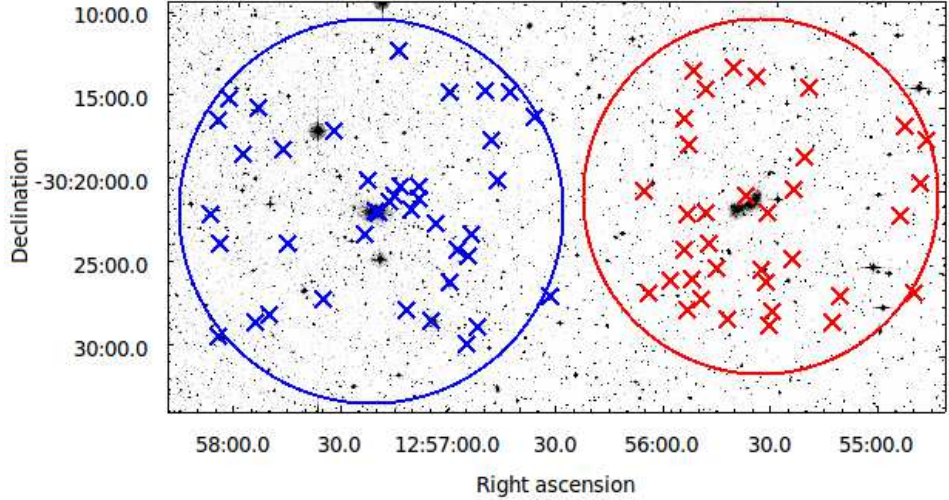}
\caption{Positions of the galaxies used for the analysis in \S\ref{sec:gal_vel_distrb_an_vir_mas} marked on the SuperCOSMOS image. The circles 
mark the 0.5R$_{200}$ radii of the clusters and the crosses mark the positions of the galaxies. Blue and red colors have been used for 
A3532 and A3530, respectively}
\label{fig:A3532_A3530_galaxy_samples}
\end{figure*}

\begin{figure*}
\centering
\includegraphics[width=3.0in]{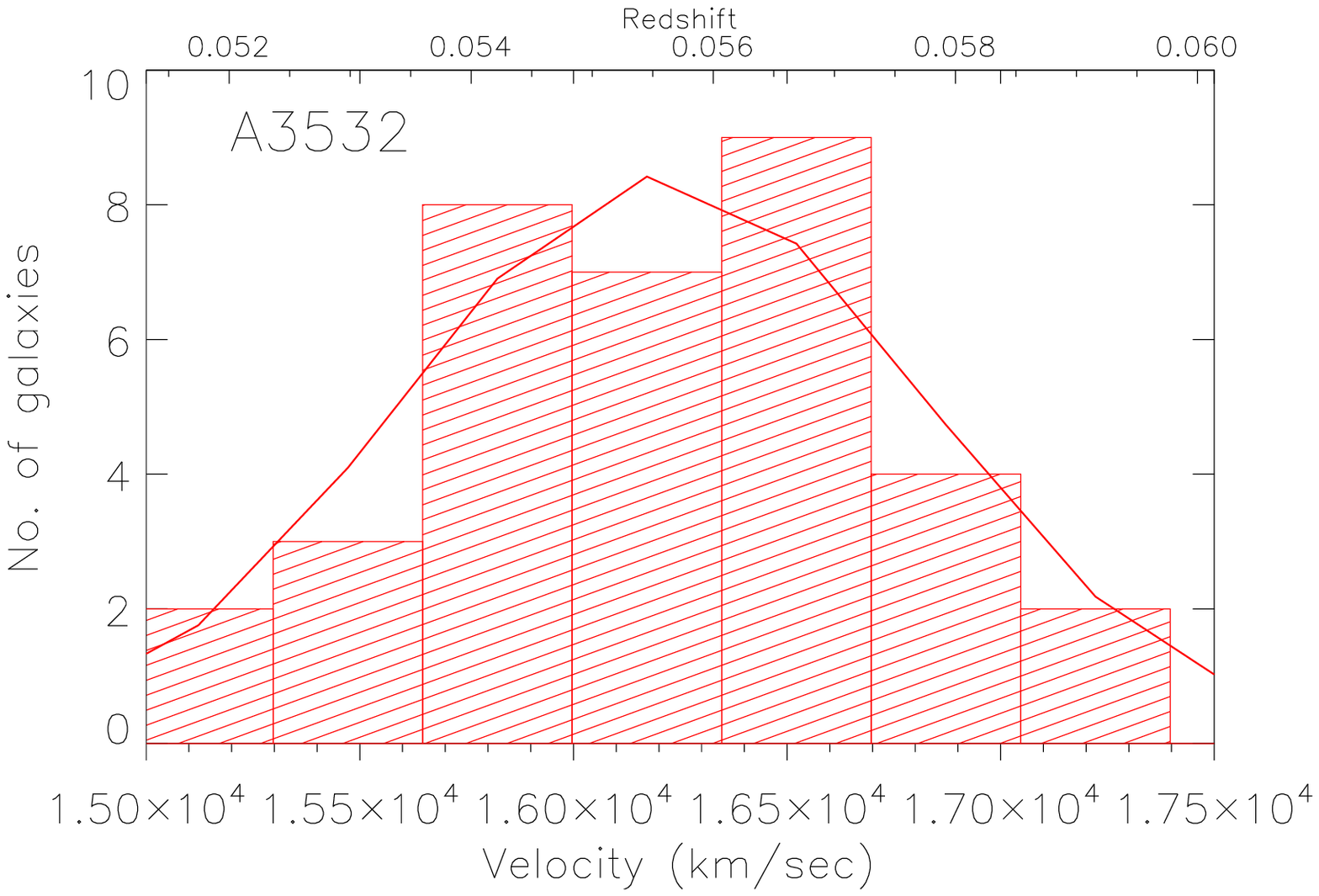}
\includegraphics[width=3.0in]{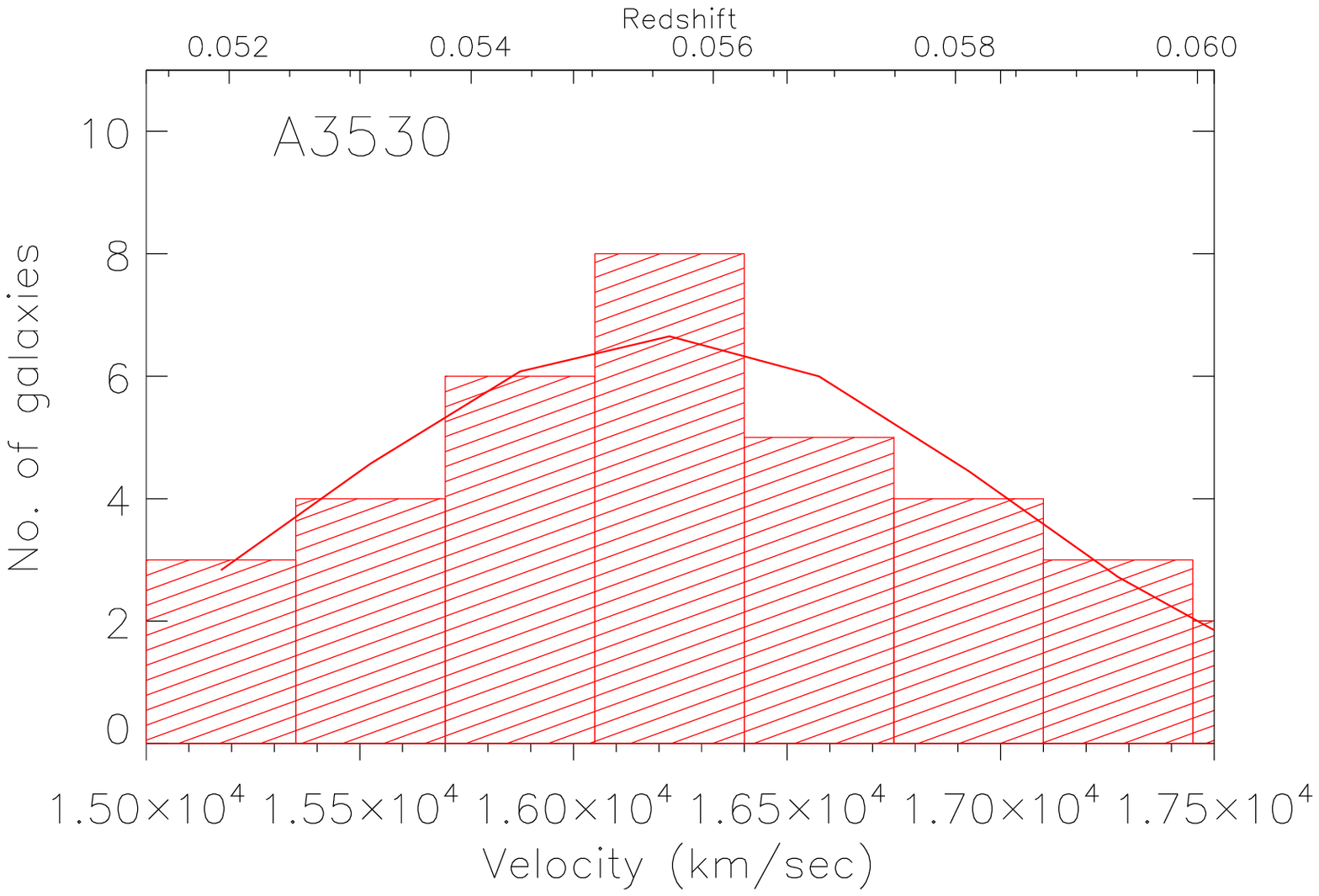}
\caption{Galaxy Velocity histograms for the clusters A3532 (left) and A3530 (right), overlaid with the Gaussian fits. The 
binsize used for both the clusters is 350 km s$^{-1}$.}
\label{fig:A3532_A3530_gal_vel_hist}
\end{figure*}

  The presence of substructures and mergers in clusters of galaxies often results in multimodal and asymmetric/Non-Gaussian velocity 
distributions. Therefore, to look for the presence of substructures and mergers in A3532 and A3530, we used the galaxy velocity samples from 
\cite{Cristiani:1987} and \cite{Bard:2001}, respectively. As \cite{Bard:2001} have given velocity information for a large number of 
galaxies located in the core of the SSC, the galaxies selected for A3530 might also have included background and foreground galaxies. 
Therefore, upper and lower velocity thresholds of 15000 and 17600 km s$^{-1}$, respectively, were applied to the sample of \cite{Bard:2001}. In 
addition, to avoid overlaps, galaxies only within 0.5 R$_{200}$ circles\footnote{The values of R$_{200}$ (radius within which the mean density 
of the cluster equals 200 times the critical density) obtained by \cite{Vulcani:2011} (1.50 Mpc for A3532 and 1.36 Mpc for A3530) have been 
used.}, centered on the X-ray surface-brightness peaks were used for this analysis (see Figure \ref{fig:A3532_A3530_galaxy_samples}), for both 
the clusters. This led to 40 galaxies with velocity information in A3532 and 35 galaxies in A3530. The histograms of galaxy velocity distributions 
of the two clusters overlaid with their Gaussian fits are shown in Figure \ref{fig:A3532_A3530_gal_vel_hist}. The  bin-size used for both 
the clusters was 350 km $\rm{s}^{-1}$. A single Gaussian can be fitted to the velocity histogram of each of the two clusters. Therefore, 
neither cluster shows the presence of substructures in its optical redshift distribution. This result is in agreement with the findings of 
\cite{Pimbblet:2008} for the cluster A3532, based on the \cite{DressShect:1988} $\delta-$test. Based on Gaussian fits, we obtain the average 
radial velocities of A3532 and A3530 as $16211\pm159$ km s$^{-1}$ and $16213\pm246$ km s$^{-1}$, respectively, which translate to average 
redshifts of 0.0556$\pm$0.0005 and 0.0556$\pm$0.0009, respectively. The result, therefore, strengthens the argument that the two clusters are 
at the same distance and much closer to each other than previously thought and, therefore, have a very high probability of tidally interacting 
with each other.

  We have also estimated the virial masses of the two clusters by using these galaxy velocity samples and the relation given by \cite{BeGeHu:1982}:
\begin{equation}
 \rm M_{\rm virial} = \frac{3 \pi}{\rm G} \sigma_{\rm r}^{2} \left\langle \frac{1}{\rm r_{\rm p}} \right\rangle^{-1} 
\end{equation}
where $\sigma_{\rm r}$ is the velocity dispersion along the line of sight and $\langle 1/\rm r_{\rm p}\rangle^{-1}$ is the harmonic 
mean projected separation between galaxy pairs. The mean velocity ($\bar{\rm v}$), velocity dispersion ($\sigma_{\rm v}$), and the
 virial masses of the clusters, thus estimated, are given in Table \ref{tab:A3532_A3530_virial_mass_param}. The underlying assumption in the 
relation used is that the galaxies included in each of the clusters are bound and their velocity dispersions are isotropic. The virial masses 
obtained for the two clusters, have large errors (specially, for A3530) and therefore, do not differ significantly. A better estimation of the 
virial masses requires more redshift data for both the clusters.

\subsection{X-ray-Radio Interaction}
\label{sec:xray_radio_interaction}
\begin{figure*}
\centering
\includegraphics[height=4.5in]{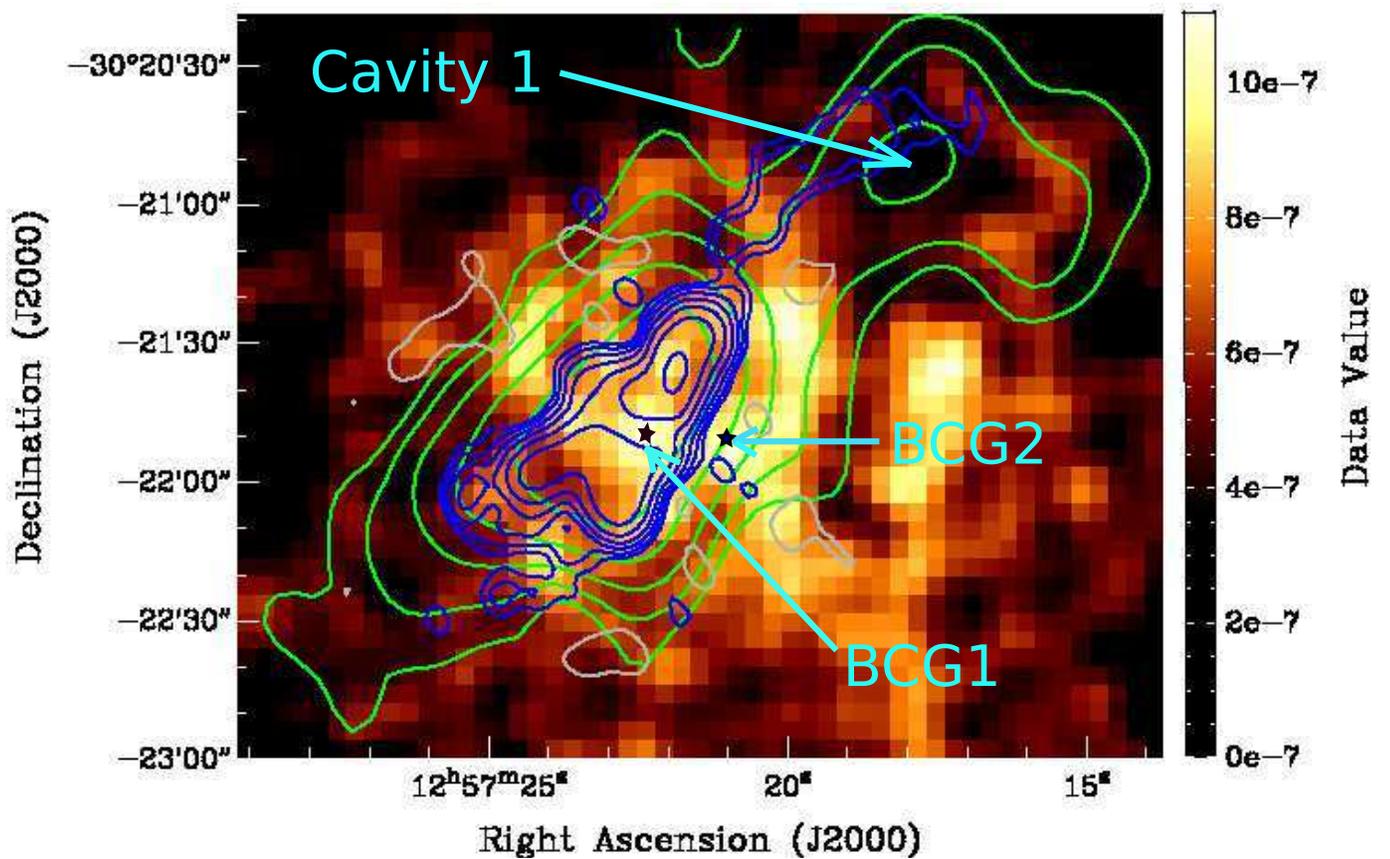}
\caption{Exposure-corrected and point-source removed \textit{Chandra} ACIS images of the central part of A3532 in the 0.3-7.0
 keV band, smoothed using a Gaussian kernel of width 4$^{\prime\prime}$, and overlaid with the GMRT 50cm contours (blue; levels same as in 
Figure \ref{fig:A3532_GMRT_50cm_WAT_source}), and the TGSS 2m radio contours (green; levels same as in 
Figure~\ref{fig:A3532_TGSS_2m_WAT_source}). Positions of the two brightest galaxies (BCG 1 and BCG 2; marked with black stars) and the 
candidate cavity (cavity 1) have been shown.}
\label{fig:Chandra_central_4arcsec_smoothedimage_ovld_radio_con}
\end{figure*}

\begin{figure*}
\centering
\includegraphics[height=4.5in,angle=270]{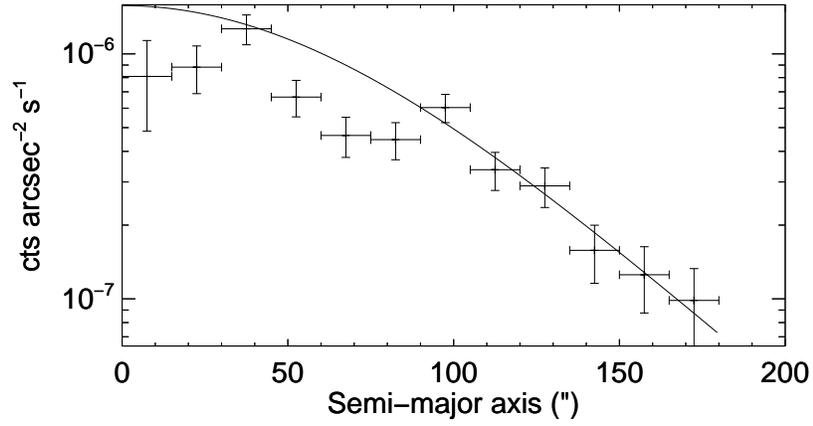}
\caption{The profile of the X-ray surface brightness along \textquoteleft{}cavity 1\textquoteright{} (see Figure 
\ref{fig:Chandra_central_4arcsec_smoothedimage_ovld_radio_con}), made by using 12 annular sectors along the direction of the cavity. A 
beta-model ($S(x)=S_{0} (1 + (x/R_{c})^{2} ) ^{-3 \beta + 0.5} $) has 
been fitted to the profile, excluding the 3 data points corresponding to the dip and the initial two points that show the central surface 
brightness fluctuations. The significance of the dip averaged over the the cavity is $\sim4\sigma$ and at the minimum is $\sim5\sigma$.}
\label{fig:large_cavity_sb_profile}
\end{figure*}

 Figure \ref{fig:Chandra_central_4arcsec_smoothedimage_ovld_radio_con} shows a moderately smoothed (Gaussian kernel width 
$\sim4^{\prime\prime}$) \textit{Chandra} image of the central part of the cluster, overlaid with the GMRT 50 cm (blue) and TGSS 2 m (green) 
contours. The image shows highly anisotropic X-ray emission with four main peaks at the centre. The brightest peak coincides with the brighter 
nucleus of the dumbbell BCG (shown as BCG 1). Another adjacent peak is seen towards its west, coinciding with the position of the second 
nucleus of the dumbbell BCG (shown as BCG 2). Two more peaks are seen at distances of about $1^{\prime}$ and $40^{\prime\prime}$, northwest 
from the center of the brightest peak. The image also shows a number of apparent cavities or depressions in the X-ray surface-brightness, both 
on large scales and small scales, and in various parts of the cluster.  However, because of the very small exposure time of the 
\textit{Chandra} observation, the detection significance of these cavities is very low. In the following analysis, we have 
focused on one \textquotedblleft{}candidate\textquotedblright{} cavity which is most prominent and visible in the Chandra images at all 
resolutions. In Figure \ref{fig:Chandra_central_4arcsec_smoothedimage_ovld_radio_con}, the 
north-western radio extension of the WAT in both sets of radio contours seems to coincide with a large scale candidate cavity 
(\textquoteleft{}cavity 1\textquoteright{}, hereafter) and in the TGSS 2m contours (green), it seems to fill the cavity completely.
 To estimate the significance of \textquoteleft{}cavity 
1\textquoteright{}, we used the X-ray surface brightness profile, made by using annular sectors along the direction of the cavity (see Figure 
\ref{fig:large_cavity_sb_profile}). \textquoteleft{}Cavity 1\textquoteright{} shows up as a significant dip 
($\sim4 \sigma$ average; $\sim5 \sigma$ at the minimum) in the X-ray surface brightness profile. It seems possible that the 
\textquoteleft{}cavity 1\textquoteright{} and the radio emission from the WAT are related to each other. Assuming \textquoteleft{}cavity 
1\textquoteright{} is indeed real, we have investigated the energy requirements of the cavity, below.

\subsubsection{Cavity Energetics}
\label{sec:cavity_thermodynamics}
The total energy required to create a cavity (E$_{cav}$) is the sum 
of the work done in expanding the cavity ($=\int$-P dV$=$PV/($\gamma$-1)) plus the energy in the cavity ($=$PV) (see Dunn \& Fabian 2004), 
where P is the pressure of the hot gas surrounding the cavity, V is the volume of the cavity, and $\gamma$ is the ratio of the specific heats 
($=$c$_{p}/$c$_{v}$). By using $\gamma=$4/3 for the relativistic jets, we obtain, E$_{cav}=\gamma PV/(\gamma-1)=4$PV (see, \cite{Birzan:2004}; 
\cite{McNamaNul2012}; \cite{Fabian:2012}). The power required by the jet to create the cavity 
(P$_{cav}$) is given by the total energy of the cavity divided by the age of the cavity. For our analysis, we have approximated the age of 
the cavity (t$_{cav}$) as the sound crossing time, which is the time taken by the sound waves to travel from the center of the AGN to the 
current location of the cavity (see \cite{Hlavacek:2012}). The pressure of the hot gas and the density (required for 
calculating the sound speed) have been estimated by using their approximate average values at the location of the cavities in the 
thermodynamic maps. To estimate the volumes of the cavities, prolate ellipsoidal shapes were assumed. The semi-major and 
semi-minor axis (R$_{1}$ and R$_{2}$) of the ellipse, the average radius R ($=\sqrt{R_{1}R_{2}}$), the pressure of the hot gas surrounding 
the cavity (P), the volume of the cavity (V), the total energy required for the cavity (E$_{cav}$), the age of the cavity (t$_{cav}$), and the 
power of the jets required (P$_{cav}$) to create the cavity, calculated for the \textquoteleft{}cavity 1\textquoteright{}, are given in 
Table \ref{tab:cavity_energetics}. Here, it should be noted that the errors associated with the shape and volume of the ellipse used to
 describe the cavity 1 were very large ($\sim$30-80\%) because of the very crude approximation and also because of the projection effects.

\begin{figure*}
\centering
\includegraphics[width=4.0in]{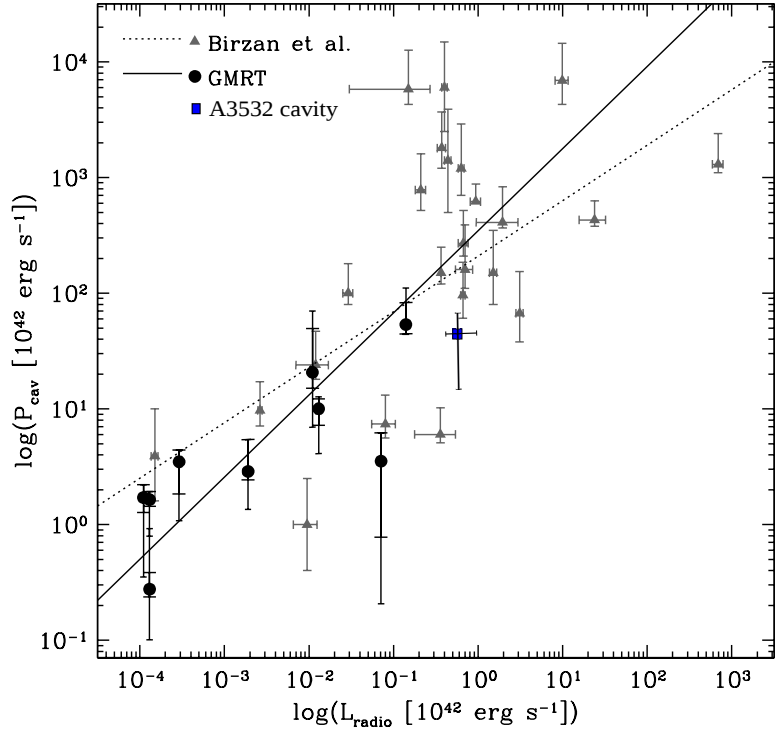}
\caption{The jet power required to create the cavity P$_{cav}$ vs. the 10 MHz-10 GHz integrated radio power L$_{radio}$, reproduced 
from OS11 (see \S\ref{sec:cavity_thermodynamics}). The 24 gray triangles and 9 black circles are corresponding to the 
cavities studied by B08 and OS11, respectively. The solid line indicates the bivariate correlated errors and intrinsic scatter (BCES) fit 
obtained by OS11 for all the data points (triangles plus circles). The dotted line indicates the relation found by B08 for only the 24 
cavities studied by them. The blue box is the point corresponding to the \textquoteleft{}cavity 1\textquoteright{}.}
\label{fig:cavity_power_radio_power}
\end{figure*}

 We have also calculated the radio power (luminosity) of the WAT source (L$_{radio}$) by integrating L$_{\nu}=4 \pi \rm 
D_{\rm L}^{2} \rm S_{\nu}$, from 10 MHz to 10 GHz.  D$_{\rm L}$ is the luminosity distance to the source, S$_{\nu} (\propto \nu^{\alpha}$) 
is the flux density at frequency $\nu$, and $\alpha$ is the radio spectral index of the WAT. The radio power so obtained, is given in Table 
\ref{tab:cavity_energetics}. L$_{radio}$ is found to be lower than P$_{cav}$, by about a factor of 80. Note that, P$_{cav}$ might have errors 
as large as 70\% due to the very crude approximation of the shape of the cavity and additional errors in the pressure of the hot gas and the 
age of the cavity. Figure \ref{fig:cavity_power_radio_power} shows the relationship between the jet power required to create the cavity and 
the radio power (integrated for the 10 MHz$\textendash$10 GHz band) reproduced from O'Sullivan et al. (2011) (OS11, hereafter). The sample 
shown in the figure is based on cavities found in the nine groups of galaxies studied by OS11 and 24 groups of galaxies studied by 
B{\^i}rzan et al.(2008) (B08, hereafter). The point corresponding to the \textquoteleft{}cavity 1\textquoteright{} (shown with a blue color) 
is found to be well within the scatter in Fig. \ref{fig:cavity_power_radio_power}. This result further supports that \textquoteleft{}cavity 
1\textquoteright{} and the radio emission from the WAT might be related to each other. The temperature in the box region at the location of 
\textquoteleft{}cavity 1\textquoteright{} in Figure \ref{fig:2d_temp_map} is found to be marginally higher than that in the immediately 
surrounding boxes. There is a possibility that the \textquoteleft{}cavity 1\textquoteright{} is formed by the energy deposited into the ICM by 
the radio jets of the WAT source in A3532 from a past central AGN outburst. In that case, P$_{cav}$ can provide an estimate for the energy 
released during that AGN outburst. From the results obtained by B{\^i}rzan et al. (2004) and Rafferty et al. (2006), we find that the typical 
observed deficit in the cooling luminosity of clusters with X-ray cavities is between $10\textendash1000\times10^{42}$ erg s$^{-1}$ and for a 
cluster with A3532$\textendash$like bolometric X-ray luminosity, it is $\sim400\times10^{42}$ erg s$^{-1}$. $P_{cav}$ for \textquoteleft{}cavity 
1\textquoteright{} is about one$\textendash$tenth of this value (see Table \ref{tab:cavity_energetics}). One can, therefore, speculate that 
only the combined effect of a few such past outbursts, could have lead to the disruption of the cool core in A3532 (see \cite{Birzan:2004}; 
\cite{Dunn:2005}; \cite{Rafferty:2006}; and \cite{Dunn:2010}). However, due to the short exposure of the Chandra observation, it is not possible 
to reliably estimate and compare the deficit in the cooling luminosity of A3532 and the total energy requirements of all the X-ray cavities 
in it.

\section{Discussion and Conclusions}
\label{sec:discussion}

 The combined image of the diffuse X-ray emission of the cluster pair A3532-A3530 shows excess X-ray emission in an 
overlapping region between the clusters (Figure \ref{fig:A3532_A3530_combined_MOS_image}). In the thermodynamic maps described in 
\S\ref{sec:box_thermodynamic_maps}, this overlapping region is found to have a significantly lower temperature and abundance than the 
clusters themselves,
 thereby nullifying the possibility of cluster scale mergers taking place between them. This observation is in agreement with the findings of 
\cite{MaudMam:2007} (see \S\ref{sec:Intro}). The results obtained by us can have two possible interpretations. In the first scenario, 
the clusters are approaching each other for the first time and are tidally interacting in their overlapping region, as a precursor to a 
possible merger in a later time.  The interaction between the two clusters has just started and, therefore, the X-ray gas in the overlapping 
region is neither very hot nor highly enriched with metals. In the second possible scenario, the individual X-ray halos of the two clusters are
 well separated from each other and the overlapping region seen between the clusters, is merely due to their chance superposition (see 
\S\ref{sec:Intro}). However, from our analysis of the galaxy velocity information available for the clusters (see 
\S\ref{sec:gal_vel_distrb_an_vir_mas}), both A3532 and A3530 seem to be at the same redshift, and therefore, we believe that the probability of
 the overlapping region being a result of a chance superposition is very low.

 A3530 shows almost constant or smoothly varying thermodynamic maps and profiles. Therefore, no significant merger activity within A3530 could 
be detected. However, there are many indications of ongoing galaxy scale mergers in the inner regions of the cluster A3532. 
These are described in the following. Firstly, the average temperature of A3532 is significantly higher than that of A3530. Secondly, the 
thermodynamic maps show high temperature regions in various parts of the cluster. Thirdly, it is seen to host a dumbbell system of BCGs at its 
center, and the brighter nucleus of the dumbbell contains a WAT radio source, which is mostly seen in merging clusters of galaxies. However, 
gravitational interaction between the galaxies of the dumbbell may also be responsible for the presence of the WAT. The overall geometry of the
 WAT, which has very closely aligned tails, may be attributed to the projection effects due to an apparently small angle between the plane of 
the WAT and our line of sight. At low frequencies, the radio emission shows an extension towards the north-west, which is either a part of the 
WAT radio emission or a separate source. The extension seems to have a steep spectrum and a rough estimate of the powerlaw spectral index is 
close to -2.

 The bolometric X-ray luminosities of the two clusters are found to be close to those of the rich clusters but with slightly higher 
average temperatures. While a high temperature in A3532 is possibly due to ongoing galaxy scale mergers in its inner regions,
 high temperature in A3530 is not clear. However, an interaction between the pair of galaxies located at the center of A3530, may have resulted
 in increasing the temperature. A deeper exposure with the \textit{Chandra} will be required to test this scenario. 

 The gas mass estimates of the clusters (\S\ref{sec:gas_mass_estimates}) show A3532 to be marginally more massive than A3530, 
while, their virial mass estimates (\S\ref{sec:gal_vel_distrb_an_vir_mas}) have large errors and do not show any significant difference. On 
comparing with the M$_{500}$ estimates of the two clusters obtained by Ettori et 
al. (1997) (see \S\ref{sec:Intro}), the virial mass obtained by us for A3530 is found to be consistent while that for A3532 is smaller. This 
discrepancy is probably because the method used by Ettori et al. and the underlying assumptions were different from our analysis and 
also because a small number of galaxies was used for the virial mass estimation, which might have led to large errors in fitting the galaxy 
velocity distribution.

 A3532 and A3530 have been classified as non cooling flow clusters in the literature (see \S\ref{sec:Intro})). The cooling 
time estimates of both clusters (see \S\ref{sec:cool_time_estimate}) are also found to be close to the Hubble time, as is observed for the 
non-cool core clusters. However, A3530 shows some other properties which are similar to the cool core clusters. A significantly low value of 
the deprojected temperature is found at its centre, where the temperature drops to about $\sim$70\% of the average value obtained for the 
cluster (see \S\ref{sec:deprojection_analysis}). The elemental abundance in A3530 is found to peak at the centre (see 
\S\ref{sec:projection_analysis} ), with a gradual decrease outwards. Negative metal abundance gradients are typical of cool core clusters 
(\cite{Johnson:2011}; \cite{DeGrandi:2001}; \cite{IrwBreg:2001}; \cite{Finoguenov:2000}). The projected global and central values of 
elemental abundance for A3530 obtained by us ($0.28^{+0.05}_{-0.04}$ and $0.48^{+0.16}_{-0.14}$ solar, respectively) are consistent 
with those of the cool core clusters ($0.37\pm0.4$ and $0.42\pm0.06$ solar, respectively;\cite{IrwBreg:2001} ). In addition, 
A3530 shows smooth, isotropic and centrally peaked profiles of X-ray surface brightness and the thermodynamic quantities (see 
\S\ref{sec:X-ray_morphology}, \S\ref{sec:projection_analysis}, \S\ref{sec:deprojection_analysis} and \S\ref{sec:box_thermodynamic_maps}), that
 are usually found in the relaxed cool core clusters. Based on all these observations the possibility of a weak cool core in A3530 can not be 
ignored. To confirm the weak cool core, deeper high-resolution X-ray observations of 
the central region of A3530 will be required. A3532, unlike A3530, shows many properties 
similar to the non cool core clusters, e.g., presence of high temperature regions at the center as well as in the other parts of the 
cluster, anisotropic variations in the X-ray surface brightness and thermodynamic maps, and no definite trend in the abundance 
profile of the cluster. A3532, therefore, seems to be a non cool core cluster where cooling flows may have been disrupted by past activity in 
the central AGN (see \cite{Fabian:2002}, \cite{McNamaNul2007}, \cite{Mittal:2009}, \cite{Baldi:2009}, \cite{Ehlert:2011}, and the discussion 
given in \S\ref{sec:cavity_thermodynamics} of this paper).

 In agreement with the findings of \cite{Pimbblet:2008}, neither cluster shows any signs of subclustering in either optical or large scale 
($\sim5^{\prime}$) X-ray emission, which is believed to be a prerequisite for an ongoing merger. However, in the high resolution 
\textit{Chandra} images of A3532 (Figures \ref{fig:Chandra_smoothedimage} and \ref{fig:Chandra_central_4arcsec_smoothedimage_ovld_radio_con}), 
we do see the presence of small scale ($20^{\prime\prime}$) substructures in its central region. The \textit{Chandra} images show a distorted 
X-ray structure and multiple peaks in the core of the cluster, which are probably a result of the stripping of the hot gas of the less bright 
dumbbell galaxy either due to tidal interaction or due to the ram pressure stripping resulting from its motion through the dense ICM, similar 
to what is observed in the dumbbell galaxies NGC 4782 and NGC 4783 (\cite{Machacek:2007}). The X-ray image also shows many candidate 
cavities in the central part of A3532 both at large scales ($\sim30^{\prime\prime}$) and small scales ($\sim10^{\prime\prime}$). 
Of these, a large scale cavity (\textquoteleft{}cavity 1\textquoteright{}) seems to coincide with the north-western extension 
of the WAT source, seen in the low frequency radio images (Figure \ref{fig:A3532_GMRT_50_cm_TGSS_MOST_843_MHz_ovld_optical}). The average 
significance of \textquoteleft{}cavity 1\textquoteright{} is found to be $\sim4$ sigma. Further, in \S\ref{sec:cavity_thermodynamics}, the 
radio luminosity of the WAT source seems to be related to the power of the jets required to create the cavity.
 If \textquoteleft{}cavity 1\textquoteright{} is indeed real then based on these results, it is possible that it is created by a buoyantly 
rising radio bubble emanating from the WAT source. The other less-significant, non-radio 
cavities, may be \textquotedblleft{}ghost\textquotedblright{} cavities, which are relics of past radio outbursts 
from the central AGN in the cluster where the radio emission has probably faded due to a stopped supply of relativistic particles from the 
nucleus (\cite{Clarke:2007}, \cite{McNamara:2001}). Deeper and high-resolution, X-ray and low frequency radio observation of A3532 are 
required for significant detection of all the cavities, to accurately study the morphological and spectral properties of the weak radio 
emission features in the cluster, and to ascertain a connection between the X-ray and radio emission from the cluster.

\section{Acknowledgements}
 The X-ray data used in this research have been obtained from the High Energy Astrophysics Science Archive Research Center (HEASARC), provided 
by NASA's Goddard Space Flight Center. We have used observations obtained with XMM-Newton, an ESA science mission with instruments and 
contributions directly funded by ESA member states and the USA (NASA), and the Chandra X-ray Observatory, managed by NASA's Marshall 
Center. We thank the Chandra and XMM helpdesk for their assistance on X-ray data analysis. Data were also obtained from the Australia 
Telescope Compact Array which is a part of the Australia Telescope National Facility, funded by the Commonwealth of Australia for operation as 
a National Facility managed by CSIRO. We have also made use of the survey data from the TGSS (http://tgss.ncra.tifr.res.in) 
with the GMRT, and from the VLSS and NVSS with the VLA. VLA is maintained by the NRAO, which is a facility of the National Science Foundation 
operated under cooperative agreement by Associated Universities, Inc. Finally, we thank the anonymous referee for his valuable 
comments and suggestions that have helped us improve many of our results.

\bibliographystyle{apj}
\bibliography{ref1}

\clearpage

\begin{sidewaystable*}
 \caption{X-ray Observations table for A3530 and A3532}
\label{tab:observation_table}
\vskip 0.5cm
\centering
{\small
\begin{tabular}{c c c c c c c c}
\hline
Cluster & Satellite & Detector &$\alpha$ (J2000)&$\delta$ (J2000)& Observation ID & Date of & Exposure \\
 & & & & & & Observation & Time \\
\hline
\hline
A3532 & \textit{XMM-Newton} & MOS1, MOS2, PN & 12 57 16.85 & -30 22 11.1 & 0030140301 & 2002 Jul 03 & 16.9 ks \\
A3532 & \textit{Chandra} & ACIS-I & 12 57 21.50 & -30 22 10. & 10745 & 2008 Nov 30 & 9.8 ks\\
A3530 & \textit{XMM-Newton} & MOS1, MOS2, PN & 12 55 35.87 & -30 19 51.4 & 0201780101 & 2004 Jan 15 & 21.9 ks \\
\hline
\end{tabular}}
\end{sidewaystable*}

\begin{sidewaystable*}
 \caption{Radio Observation table for A3532 WAT.}
\label{tab:radio_observation_table}
\vskip 0.5cm
\centering
  \begin{threeparttable}
{\small
\begin{tabular}{c c c c c c c c c}
\hline
Observation & Wavelength &Beam Size &Position & Array &Date of&RMS & Flux Density & Reference\\
 & m & & Angle &Configuration& Observation & mJy/beam & Jy & \\
\hline
\hline
VLA & 0.06 &$13^{\prime\prime}.37\times 9^{\prime\prime}.66$ & -54.49 & VLA-DnC & 1991 Feb 05 & 0.035 & 0.36$\pm$0.02 & P\\
ATCA & 0.13 &$6^{\prime\prime}.28\times 4^{\prime\prime}.08$ & -0.26 & 6C & 1994 Mar 15 & 0.3 & 0.69$\pm$0.03 & P\\
VLA & 0.20 &$4^{\prime\prime}.87\times 2^{\prime\prime}.91$ & -52.73 & VLA-BnA & 1990 Jul 09 & 0.12 & 1.05$\pm$0.05 & P\\
VLA (NVSS) & 0.21 &$45^{\prime\prime}\times 45^{\prime\prime}$ & 0 & -- & -- & 0.48 & 1.16$\pm$0.04 & 1\\
ATCA & 0.22 &$10^{\prime\prime}.72\times 6^{\prime\prime}.45$ & 0.02 & 6C & 1994 Mar 15 & 0.4 & 1.06$\pm$0.05 & P\\
MOST (SUMSS) & 0.36 &$85^{\prime\prime}\times 45^{\prime\prime}$ & 0.0 & -- & -- & 1.8 & 1.78$\pm$0.05 & 2 \\
GMRT & 0.50 & $7^{\prime\prime}.45\times 4^{\prime\prime}.62$ & 28.90 & & 2004 Apr 04 & 0.13 & 1.9$\pm$0.3 & P\\
Molonglo Cross (MRC) & 0.74 &$2.87^{\prime}\times 2.62^{\prime}$ & 0.0 & -- & -- & 50 & 3.03$\pm$0.08 & 3\\
GMRT (TGSS) & 2.0 &$24^{\prime\prime}\times 15^{\prime\prime}$ & 30 & -- & -- & 23 & 8.7$\pm$2.2 & 4\\
VLA (VLSS) & 4.0 &$58.8^{\prime\prime}\times 36.2^{\prime\prime}$ & -47.9 & -- & -- & 102 & 11.7$\pm$1.2 & 5\\
\hline
\end{tabular}}
     \begin{tablenotes}
\footnotesize 
       \item[] references:
       \item[] P: Present paper, 1: Condon et al.(1998), 2: Mauch et al. (2003), 3: Large et al. (1981), 4: TGSS Data Products, 
5: Cohen et al. (2007).

       \item[Notes] (a) Errors of $\sim$5\% have been assumed for both the ATCA observations, the VLA 1.49 and 4.86 GHz 
observations. For the GMRT TGSS and 608 MHz observations errors of $\sim$25\% and $\sim$15\% have been assumed, respectively. (b) For the ATCA 
observations, the pointing center (R.A. (J2000), Dec. (J2000)) was 12$^{h}$ 57$^{m}$ 13.0$^{s}$, -30$^{\circ}$ 23$^{\prime}$ 
44$^{\prime\prime}$ (J2000), the bandwidth was 128 MHz, and the total integration time was approximately 2 hours. For the GMRT observations, 
the pointing center for the observation was 12$^{h}$ 56$^{m}$ 30$^{s}$, -30$^{\circ}$ 30$^{\prime}$ (J2000), the bandwidth was 16 MHz, and the 
total integration time was about 2 hour 12 minutes. For the VLA L-band and C-band observations, the pointing centers were 12$^{h}$ 57$^{m}$ 
20.85$^{s}$, -30$^{\circ}$ 21$^{\prime}$ 51.64$^{\prime\prime}$ and 12$^{h}$ 57$^{m}$ 21.99$^{s}$, -30$^{\circ}$ 21$^{\prime}$ 
47.42$^{\prime\prime}$, and the total integration times were $\sim$25 minutes and  $\sim$15 minutes, respectively. The bandwidth for both 
observations was 50 MHz.
     \end{tablenotes}
  \end{threeparttable}
\end{sidewaystable*}

\begin{sidewaystable*}
 \caption{Best fit parameters obtained from the spectral analysis done using both \textit{XMM-Newton} and \textit{Chandra} data for the 
cluster A3532, and using only \textit{XMM-Newton} data for the cluster A3530 and the overlapping region (OR) between the clusters 
(\S\ref{sec:Total-Spec-Analy}). X-ray luminosities of the two clusters derived in \S\ref{sec:luminosity_estimates} are given. Results from 
Ikebe et al. (2002) based on \textit{ASCA} data are also given for comparison.}
\label{tab:total_regions_spectral_results}
\vskip 0.3cm
\centering
  \begin{threeparttable}
\begin{tabular}{c c c c c c c c}
 \hline
\hline
Region&Satellite&kT&Abundance & \textit{apec} norm.&L$_{\rm x}$&L$_{\rm x}$\footnotemark[1]&$(\chi^{2}_{\nu})_{min}$ (DOF)\\
 & & & & &(0.5-8.0 keV)&(0.1-2.4 keV)& \\
 & & (keV)&(Rel. to solar)& ($10^{-3}$ cm$^{-5}$) &($10^{43}$ergs s$^{-1}$)&($10^{43}$ergs s$^{-1}$)& \\
\hline
A3532 & \textit{XMM-Newton}  & $4.8\pm0.2        $& $0.36\pm0.08$          & $16.2\pm0.4$ & $14.0\pm0.3$ & $10.2\pm0.1$ & 1.23 (364)\\
A3532 & \textit{Chandra}     & $5.2^{+0.6}_{-0.6}$& $0.4^{+0.3}_{-0.3}$    & $14.6\pm1.1$ & $13.2\pm0.6$ & -- & 0.97 (47)\\
A3532 & \textit{ASCA\footnotemark[2]}  & $4.4\pm{0.2}$& --  & -- & -- &$12.0\pm0.4$ & --\\
A3530 & \textit{XMM-Newton}  & $3.5\pm0.1$        & $0.28^{+0.05}_{-0.04}$ & $9.9\pm0.2$  & $7.6\pm0.1$  & $4.9\pm0.1$ & 1.13 (705)\\
A3530 & \textit{ASCA\footnotemark[2]}  & $4.1\pm0.3$& --  & -- & -- & $6.6\pm0.3$ & --\\
OR & \textit{XMM-Newton}  & $2.1^{+0.3}_{-0.2}$& $0.09^{+0.09}_{-0.07}$ & $1.6\pm0.2$  & $0.81\pm0.05$  & -- & 1.15 (76)\\[1ex]
\hline
\end{tabular}
     \begin{tablenotes}
\footnotesize 
       \item[1] The column shows the 0.1-2.4 keV luminosities of A3532 and A3530 obtained by us using \textit{XMM-Newton} data, for a 
comparison with the results of Ikebe et al. (2002). The values of hydrogen column density and the redshifts have been frozen to the values 
used by Ikebe et al., although their values, which were used for the rest of the analyses, do not change the results significantly.
       \item[2] The rows show the X-ray temperatures of the hot components and the 0.1-2.4 keV luminosities of A3532 and A3530, obtained by 
Ikebe et al. (2002) by fitting 2-T thermal plasma model to the ASCA spectra of the clusters. The luminosities have been scaled for the 
currently used value of the Hubble Constant (=70 km s$^{-1}$ Mpc$^{-1}$).
       \item[] \underline{Notes}: The spectra for each region is fitted with a single-temp \textit{apec} model for a fixed 
Galactic absorption. For the \textit{XMM-Newton} spectral analysis, the residual soft proton contamination and the instrumental Al line at 
1.49 keV have been modeled by adding powerlaws and Gaussians respectively (separately for MOS1, MOS2 and PN) to the models. Best fit 
values for the temperature (kT), elemental abundance relative to the solar values, normalization of the \textit{apec} model, X-ray 
luminosity (L$_{\rm x}$), and minimum reduced $\chi^{2}_{\nu}$ are given along with the degrees of freedom (DOF). For the \textit{XMM-Newton} 
spectral analysis of the cluster A3532, only MOS1 and MOS2 data have been used while for the cluster A3530 and the overlapping region, PN data 
has also been used.\\
All errors are quoted at 90\% confidence level based on $\chi^{2}_{min}$+2.71.
     \end{tablenotes}
  \end{threeparttable}
\end{sidewaystable*}
\clearpage

\begin{table*}
 \caption{Results of $\beta$-model fitting of the surface-brightness profiles of the clusters A3532 and A3530 and their 
bolometric X-ray luminosities.}
\label{tab:bol_xray_luminosity}
\vskip 0.5cm
\centering
  \begin{threeparttable}
\begin{tabular}{c c c c  c}
\hline
\hline 
Cluster & $\beta$               & $r_{c}$ & $\rm F^{\rm b}_{\rm X}$ & $\rm L^{\rm b}_{\rm X}$ \\
 &  & ($10^{-2}$ Mpc) & ($10^{-11}$ erg cm$^{-2}$ s$^{-1}$) & ($10^{44}$ erg s$^{-1}$) \\
\hline
A3532   & $0.41\pm0.01$  & $8.1\pm0.5$ & $3.4\pm0.3$ & $2.5\pm0.2$ \\
A3530   & $0.47\pm0.01$  & $9.7\pm0.5$ & $1.3\pm0.2$ & $0.9\pm0.1$ \\
\hline
\end{tabular}
     \begin{tablenotes}
\footnotesize
       \item[a] Errors are quoted at 68\% confidence level (1$\sigma$) based on $\chi^{2}_{min}$+1.00.
       \item[b] $\rm F_{\rm x}(0)$ $=$ Bolometric X-ray flux, $\beta$ $=$ $-1/3$(slope-0.5) and $\rm r_{\rm c}$ $=$ core radius.
     \end{tablenotes}
  \end{threeparttable}
\end{table*}

\begin{table*}
 \caption{Best-fit parameters obtained from the spectral analysis of eight circular annuli in the cluster A3532 using \textit{XMM-Newton} MOS1 
and MOS2 data and of five circular annuli using \textit{Chandra} data.}
\label{tab:A3532_projected_annuli_spectral_results}
\vskip 0.5cm
\centering
  \begin{threeparttable}
\begin{tabular}{c c c c c c}
\hline
\hline
Data & Annulus Number& kT & n$_{\rm e}$ & P & S \\
 & & (keV) & ($10^{-4}$ cm$^{-3}$) & ($10^{-12}$ dyn cm$^{-2}$) & (keV cm$^{2}$) \\
\hline
\textit{XMM-Newton} & 1 & $4.1\pm0.4$         & $43.2\pm0.7$ & $28.3^{+3.2}_{-5.5}$ & $155\pm17$ \\
           & 2 & $5.2\pm0.4$         & $21.6\pm0.3$ & $18.0^{+1.6}_{-2.0}$ & $311\pm26$ \\
           & 3 & $4.8\pm0.4$         & $13.3\pm0.2$ & $10.2^{+1.0}_{-1.4}$ & $398\pm36$ \\
           & 4 & $5.1^{+0.6}_{-0.5}$ &  $8.9\pm0.1$ &  $7.3^{+1.0}_{-1.2}$ & $550^{+69}_{-59}$ \\
           & 5 & $4.6^{+0.6}_{-0.5}$ &  $6.4\pm0.1$ &  $4.7^{+0.7}_{-0.8}$ & $619^{+87}_{-74}$ \\
           & 6 & $4.3^{+0.7}_{-0.5}$ &  $4.7\pm0.1$ &  $3.2^{+0.6}_{-0.8}$ & $711^{+125}_{-92}$ \\
           & 7 & $4.1^{+0.9}_{-0.6}$ &  $3.6\pm0.1$ &  $2.4^{+0.6}_{-0.8}$ & $810^{+192}_{-133}$ \\
           & 8 & $3.3^{+0.9}_{-0.6}$ &  $2.8\pm0.1$ &  $1.5^{+0.5}_{-0.4}$ & $765^{+225}_{-156}$ \\
\hline
\textit{Chandra} & 1 & $5.1^{+0.8}_{-0.7}$ &  $42.8\pm0.9$ & $35.0^{+6.2}_{-5.5}$ & $193^{+33}_{-29}$ \\
        & 2 & $4.8^{+0.6}_{-0.5}$ &  $21.0\pm0.4$ & $16.1^{+2.3}_{-2.0}$ & $292^{+40}_{-34}$ \\
        & 3 & $4.8^{+0.7}_{-0.6}$ &  $12.2\pm0.2$ &  $9.4^{+1.5}_{-1.3}$ & $420^{+67}_{-58}$ \\
        & 4 & $5.1^{+1.1}_{-0.8}$ &   $8.3\pm0.2$ &  $6.8^{+1.6}_{-1.2}$ & $576^{+133}_{-99}$ \\
        & 5 & $3.7^{+1.1}_{-0.8}$ &   $5.5\pm0.2$ &  $3.2^{+1.1}_{-0.8}$ & $554^{+177}_{-132}$ \\
\hline
\end{tabular}
    \begin{tablenotes}
\footnotesize
       \item[Notes.] The spectra for all the annuli were fitted using the model \textit{wabs*apec} for a fixed value of Galactic 
absorption and with elemental abundances frozen to 0.36 times the solar value. For \textit{XMM-Newton} spectra the residual soft proton 
contamination and the instrumental Al line at 1.49 keV have been modeled by adding powerlaws and Gaussians respectively (separately 
for MOS1 and MOS2) to the models. Annulus number represents the position of the annulus from the innermost to the outermost annuli in 
increasing order. Values of temperature (kT), electron density (n$_{\rm e}$), pressure (P), and entropy (S) are listed.\\
All errors are quoted at 90\% confidence level based on $\chi^{2}_{min}$+2.71.
     \end{tablenotes}
  \end{threeparttable}
\end{table*}

\begin{table*}
 \caption{Best fit parameters obtained from the spectral analysis of eight circular annuli in the cluster A3530 using 
\textit{XMM-Newton} data.}
\label{tab:A3530_projected_annuli_spectral_results}
\vskip 0.5cm
\centering
  \begin{threeparttable}
\begin{tabular}{c c c c c c}
\hline
\hline
 Annulus Number& kT & Abundance & n$_{\rm e}$ & P & S \\
 & (keV) & (relative to solar) & ($10^{-4}$ cm$^{-3}$) & ($10^{-12}$ dyn cm$^{-2}$) & (keV cm$^{2}$) \\
\hline
 1 & $3.7\pm0.2$ & $0.5^{+0.2}_{-0.1}$ & $31.4\pm0.8$ & $18.6\pm1.5$ & $172\pm12$ \\
 2 & $4.2\pm0.2$         & $0.3\pm0.1$ & $16.1\pm0.3$ & $10.8\pm0.7$         & $306\pm18$ \\
 3 & $3.6\pm0.2$         & $0.2\pm0.1$ &  $9.8\pm0.2$ &  $5.7\pm0.4$         & $364\pm25$ \\
 4 & $3.1\pm0.2$         & $0.3\pm0.1$ &  $6.6\pm0.2$ &  $3.3\pm0.3$         & $411\pm33$ \\
 5 & $2.9\pm0.3$         & $0.1\pm0.1$ &  $5.0\pm0.1$ &  $2.3\pm0.3$         & $460\pm56$ \\
 6 & $2.6^{+0.4}_{-0.2}$ & $0.2\pm0.1$ &  $3.8\pm0.2$ &  $1.6^{+0.3}_{-0.2}$ & $491^{+89}_{-51}$ \\
 7 & $1.9\pm0.2$         & $0.1\pm0.1$ &  $3.2\pm0.1$ &  $1.0\pm0.1$         & $407\pm55$ \\
 8 & $1.6\pm0.2$         & $0.04^{+0.06}_{-0.04}$ &  $2.6\pm0.1$ &  $0.7\pm0.1$         & $397\pm65$ \\
\hline
\end{tabular}
    \begin{tablenotes}
\footnotesize
       \item[Notes.] The spectra for all the annuli were fitted using the model \textit{wabs*apec} for a fixed value of Galactic 
absorption.\\ The residual soft proton contamination and the instrumental Al line at 1.49 keV have been modeled by adding 
powerlaws and Gaussians respectively (separately for MOS1, MOS2 and PN) to the models. Annulus number 
represents the position of the annulus from the innermost to the outermost annuli in increasing order. Values of
 temperature (kT), abundance, electron density (n$_{\rm e}$), pressure (P), and entropy (S) are listed.\\
 All errors are quoted at 90\% confidence level based on $\chi^{2}_{min}$+2.71.
     \end{tablenotes}
  \end{threeparttable}
\end{table*}
\clearpage

\begin{table*}
 \caption{Best fit parameters obtained from the deprojected spectral analysis of eight circular annuli in the cluster A3532 
using \textit{XMM-Newton} MOS1 and MOS2 data and of five circular annuli using \textit{Chandra} data.}
\label{tab:A3532_deprojected_annuli_spectral_results}
\vskip 0.5cm
\centering
  \begin{threeparttable}
\begin{tabular}{c c c c c c}
\hline
\hline
Data & Annulus Number& kT & n$_{\rm e}$ & P & S \\
 & & (keV) & ($10^{-4}$ cm$^{-3}$) & ($10^{-12}$ dyn cm$^{-2}$) & (keV cm$^{2}$) \\
\hline
\textit{XMM-Newton} & 1 & $3.7^{+0.9}_{-0.6}$ & $30.3\pm1.1$ & $17.9^{+5.0}_{-3.6}$ & $177^{+47}_{-33}$ \\
           & 2 & $4.0^{+0.9}_{-0.7}$ & $15.8\pm0.6$ & $10.1^{+2.6}_{-2.1}$ & $295^{+74}_{-59}$ \\
           & 3 & $4.4^{+1.0}_{-0.7}$ & $11.8\pm0.4$ &  $8.3^{+2.1}_{-1.6}$ & $395^{+98}_{-71}$ \\
           & 4 & $5.0^{+1.8}_{-1.1}$ &  $8.2\pm0.3$ &  $6.5^{+2.6}_{-1.7}$ & $572^{+221}_{-141}$ \\
           & 5 & $5.0^{+1.8}_{-1.5}$ &  $6.6\pm0.4$ &  $5.3^{+2.2}_{-1.9}$ & $661^{+265}_{-225}$ \\
           & 6 & $4.3^{+1.9}_{-1.0}$ &  $5.0\pm0.3$ &  $3.4^{+1.7}_{-1.0}$ & $682^{+325}_{-183}$ \\
           & 7 & $3.4^{+2.5}_{-1.2}$ &  $3.3\pm0.3$ &  $1.8^{+1.5}_{-0.8}$ & $713^{+573}_{-301}$ \\
           & 8 & $2.5^{+0.5}_{-0.4}$ &  $4.5\pm0.2$ &  $1.8^{+0.4}_{-0.4}$ & $426^{+96}_{-79}$ \\
\hline
\textit{Chandra} & 1 & $5.4^{+2.2}_{-1.4}$ &  $29.5\pm1.3$ & $25.5^{+11.2}_{-7.7}$& $263^{+115}_{-76}$ \\
        & 2 & $4.8^{+1.4}_{-1.0}$ &  $17.8\pm0.7$ & $13.7^{+4.5}_{-3.4}$ & $327^{+104}_{-76}$ \\
        & 3 & $4.5^{+1.8}_{-1.1}$ &  $11.0\pm0.5$ &  $7.9^{+3.5}_{-2.3}$ & $423^{+183}_{-117}$ \\
        & 4 & $6.1^{+4.7}_{-1.8}$ &   $8.5\pm0.4$ &  $8.3^{+6.8}_{-2.9}$ & $680^{+547}_{-224}$ \\
        & 5 & $3.9^{+1.1}_{-0.7}$ &   $8.2\pm0.3$ &  $5.1^{+1.6}_{-1.1}$ & $445^{+135}_{-90}$ \\
\hline
\end{tabular}
    \begin{tablenotes}
\footnotesize
       \item[Notes.]  The spectra for all the annuli were fitted using the model \textit{wabs*apec} for a fixed value of Galactic 
absorption and with elemental abundances frozen to 0.36 times the solar value. For \textit{XMM-Newton} spectra the residual soft proton 
contamination and the instrumental Al line at 1.49 keV have been modeled by adding powerlaws and Gaussians respectively to the models.
 Annulus number represents the position of the annulus from the innermost to the outermost annuli in increasing order. Values of 
temperature (kT), electron density (n$_{\rm e}$), pressure (P), and entropy (S) are listed.\\
All errors are quoted at 90\% confidence level based on $\chi^{2}_{min}$+2.71.
     \end{tablenotes}
  \end{threeparttable}
\end{table*}

\begin{table*}
 \caption{Best fit parameters obtained from the deprojected spectral analysis of eight circular annuli in the cluster A3530
 using \textit{XMM-Newton} data.}
\label{tab:A3530_deprojected_annuli_spectral_results}
\vskip 0.5cm
\centering
  \begin{threeparttable}
\begin{tabular}{c c c c c}
\hline
\hline
 Annulus Number& kT & n$_{\rm e}$ & P & S \\
 & (keV) & ($10^{-4}$ cm$^{-3}$) & ($10^{-12}$ dyn cm$^{-2}$) & (keV cm$^{2}$) \\
\hline
 1 & $2.5^{+0.2}_{-0.1}$ & $21.4\pm0.7$ & $8.6^{+1.0}_{-0.6}$ & $150^{+15}_{-9}$ \\
 2 & $3.7\pm0.6$         & $12.4\pm0.3$ & $7.3\pm1.4$          & $321\pm57$ \\
 3 & $3.9^{+0.8}_{-0.5}$ &  $8.4\pm0.2$ &  $5.2^{+1.2}_{-0.8}$ & $438^{+97}_{-64}$ \\
 4 & $2.7^{+0.7}_{-0.4}$ &  $6.0\pm0.2$ &  $2.6^{+0.8}_{-0.5}$ & $380^{+107}_{-65}$ \\
 5 & $2.7^{+0.4}_{-0.3}$ &  $4.2\pm0.2$ &  $1.8^{+0.4}_{-0.3}$ & $479^{+85}_{-67}$ \\
 6 & $3.2^{+0.6}_{-0.5}$ &  $4.0\pm0.2$ &  $2.1^{+0.5}_{-0.4}$ & $586^{+127}_{-108}$ \\
 7 & $2.7^{+0.5}_{-0.4}$ &  $2.9\pm0.2$ &  $1.3\pm0.3$         & $611^{+139}_{-116}$ \\
 8 & $2.2^{+0.4}_{-0.3}$ &  $3.9\pm0.1$ &  $1.4^{+0.3}_{-0.2}$ & $412^{+82}_{-63}$ \\

\hline
\end{tabular}
    \begin{tablenotes}
\footnotesize
       \item[Notes.] The spectra for all the annuli were fitted using the model \textit{wabs*apec} for a fixed value of 
Galactic absorption and with elemental abundances frozen to 0.28 times the solar value. The residual soft proton 
contamination and the instrumental Al line at 1.49 keV have been modeled by adding powerlaws and Gaussians respectively 
to the models. Annulus number represents the position of the annulus from the innermost to the outermost annuli in 
increasing order. Values of temperature (kT), electron density (n$_{\rm e}$), pressure (P), and entropy (S) are listed.\\
 All errors are quoted at 90\% confidence level based on $\chi^{2}_{min}$+2.71.
     \end{tablenotes}
  \end{threeparttable}
\end{table*}
\clearpage

\begin{sidewaystable*}
 \caption{Mass of hot gas for the clusters A3532 and A3530 obtained by fitting $\beta$-models to the projected and deprojected density 
profiles obtained in \S\ref{sec:projection_analysis} and \S\ref{sec:deprojection_analysis}, respectively.}
\label{tab:gas_mass_estimate}
\vskip 0.5cm
\centering
  \begin{threeparttable}
\begin{tabular}{c c c c c c c c}
\hline
\hline
Spectral & Cluster & Data & $\beta$               & $r_{c}$ & $\rho_{0}$ & r & $\rm M_{\rm gas } (r)$ \\
Analysis & & &  & (kpc) & ($10^{13}$ $\rm M_{\odot}$ Mpc$^{-3}$) & (Mpc) & ($10^{13}$ $\rm M_{\odot}$) \\
\hline
Projected   & A3532 & XMM-Newton & $0.68\pm0.01$           & $77\pm3$  & $10.3\pm0.1$ & 0.5 & $0.70\pm0.04$\\
            &       &            &                         &           &              & 1.0 & $2.3\pm0.2$ \\
            &       & Chandra    & $0.68\pm0.03$           & $71\pm1$  & $10.6\pm0.2$ & 0.5 & $0.6\pm0.1$ \\
            &       &            &                         &           &              & 1.0 & $2.1\pm0.3$ \\
            & A3530 & XMM-Newton & $0.64\pm0.01$           & $67\pm2$  & $8.0\pm0.1$  & 0.5 & $0.53\pm0.03$ \\
            &       &            &                         &           &              & 1.0 & $1.8\pm0.1$ \\
\hline
Deprojected & A3532 & XMM-Newton & $0.54\pm0.03$           & $71\pm9$  & $7.1\pm0.2$  & 0.5 & $0.7\pm0.1$ \\
            &       &            &                         &           &              & 1.0 & $2.6\pm0.5$ \\
            &       & Chandra    & $0.52\pm0.08$           & $72\pm2$  & $7.1\pm0.3$ & 0.5 & $0.7\pm0.2$ \\
            &       &            &                         &           &             & 1.0 & $2.8\pm1.2$ \\
            & A3530 & XMM-Newton & $0.54\pm0.02$           & $73\pm8$  & $5.1\pm0.1$ & 0.5 & $0.5\pm0.1$ \\
            &       &            &                         &           &             & 1.0 & $2.0\pm0.3$ \\
\hline
\end{tabular}
     \begin{tablenotes}
\footnotesize
       \item[a] Errors are quoted at 68\% confidence level (1$\sigma$) based on $\chi^{2}_{min}$+1.00.
     \end{tablenotes}
  \end{threeparttable}
\end{sidewaystable*}

\begin{table*}
 \caption{The values of virial mass and the parameters used (derived in \S\ref{sec:gal_vel_distrb_an_vir_mas}), for A3532 and A3530}
\label{tab:A3532_A3530_virial_mass_param}
\vskip 0.5cm
\centering
  \begin{threeparttable}
\begin{tabular}{c c c c c}
\hline
\hline
Cluster & No. of galaxies & $\bar{\rm v}$ & $\sigma_{\rm v}$ & $\rm M_{\rm virial}$\\
 &  & (km s$^{-1}$) & (km s$^{-1}$) & ($10^{14}$ $\rm M_{\odot}$)\\
\hline
A3532 & 40 & $16211\pm157$ & $615\pm159$ & $3.4\pm1.8$ \\
A3530 & 35 & $16213\pm246$ & $794\pm286$ & $5.5\pm4.0$ \\[1ex]

\hline
\end{tabular}
     \begin{tablenotes}
\footnotesize
       \item[a] Errors are quoted at 90\% confidence level based on $\chi^{2}_{min}$+2.71.
     \end{tablenotes}
  \end{threeparttable}
\end{table*}

\begin{table*}
 \caption{Energetics of the candidate \textquoteleft{}cavity 1\textquoteright{} (see \S\ref{sec:xray_radio_interaction}). R$_{1}$ 
and R$_{2}$ and R represent the semi-major axis, the semi-minor axis and the average radius ($=\sqrt{R_{1}R_{2}}$) of the approximate ellipses 
describing the cavities, respectively. P is the pressure of the hot gas surrounding the cavities, V is the volume of the (prolate) ellipsoidal 
cavities, E$_{cav}$ is the total energy required to create the cavities, t$_{cav}$ is the cavity age, P$_{cav}$ is the jet power required, 
and L$_{radio}$ is the 10 MHz-10 GHz integrated radio power of the WAT source. Details are given in \S\ref{sec:cavity_thermodynamics}.}
\label{tab:cavity_energetics}
\vskip 0.5cm
\centering
\begin{tabular}{c c c c c c c c c}
\hline
\hline
R$_{1}$ & R$_{2}$ & R & P & V & E$_{cav}$ & t$_{cav}$ & P$_{cav}$ & L$_{radio}$ \\
(kpc) & (kpc) & (kpc) & ($10^{-11}$ erg cm$^{-3}$) & ($10^{69}$ cm$^{3}$) &($10^{59}$ erg) &(10$^{7}$ yr) &(10$^{43}$ erg s$^{-1}$) 
& (10$^{43}$ erg s$^{-1}$)\\
\hline
23.9 & 21.3  & 22.6 & 2.2 &  1.2 & 1.2  & 7.5 & 4.6 & $0.06^{+0.03}_{-0.02}$\\
\hline
\end{tabular}
\end{table*}


\clearpage

\clearpage

\clearpage  

\clearpage  

\clearpage

\clearpage

\clearpage

\clearpage

\clearpage

\clearpage

\end{document}